\documentclass{LMCS}

\def\dOi{10(3:16)2014}
\lmcsheading%
{\dOi}
{1--36}
{}
{}
{Oct.~23, 2012}
{Sep.~23, 2014}
{}

\ACMCCS{[{\bf Theory of computation}]: Logic---Proof theory; Linear logic}


\usepackage{hyperref}
\usepackage[all]{xypic}
\usepackage{amssymb}
\input{bellin-macros.sty}
\input{bussproofs.sty}

\begin{document}

\title[Co-Intuitionistic Linear Logic]{Categorical Proof Theory of Co-Intuitionistic Linear Logic}

\author[Gianluigi Bellin]{Gianluigi Bellin}	
\address{Dipartimento di Informatica, Universit\`a di Verona, Strada Le Grazie, 37134 Verona Italy}	
\email{gianluigi.bellin@univr.it}  





\begin{revision}
  This is a revised and corrected version of the article originally
  published on September 10, 2014.
\end{revision}

\begin{abstract}
  \noindent 
  To provide a categorical semantics for co-intuitionistic logic, one
  has to face the fact, noted by Tristan Crolard, that the definition
  of co-exponents as adjuncts of co-products does not work in the
  category {\bf Set}, where co-products are disjoint unions. Following
  the familiar construction of models of intuitionistic linear logic
  with exponential ``{\bf !}'', we build models of co-intuitionistic
  logic in symmetric monoidal closed categories with additional
  structure, using a variant of Crolard's term assignment to
  co-intuitionistic logic in the construction of a free category.
\end{abstract}

\maketitle

\section*{Preface}\label{preface}
%
%

This paper sketches a categorical semantics for co-intuitionistic logic, advancing a line of 
proof-theoretic research developed in \cite{Bellin12,BellinLAM,Bellin05,Bellin04,BellinMenti}. 
Co-intuitionistic logic, also called {\em dual-intuitionistic} \cite{Gore,Tranchini,Wansing}, 
may be superficially regarded as completely determined by the duality, as in its lattice-theoretic 
semantics. A {\em co-Heyting algebra} is a (distributive) lattice $\mathcal{C}$ such 
that its opposite $\mathcal{C}^{op}$ is a Heyting algebra. 
In a Heyting algebra {\em implication} $B \rightarrow A$ is defined as the right adjoint of {\em meet}, 
so in a co-Heyting algebra $\mathcal{C}$ {\em co-implication} (or {\em subtraction}) 
 $A \smallsetminus B$ is defined as the left adjoint of {\em join}:
$$
{{C\wedge B\ \leq\ A} \over {C\ \leq\ B\rightarrow A}}\hskip1in %
{{A\ \leq\ B\vee C} \over {A\smallsetminus B\ \leq\ C}}
$$
A {\em bi-Heyting} algebra is a lattice that has both the structure of a Heyting 
and of a co-Heyting algebra. The logic of bi-Heyting algebras was introduced 
by Cecylia Rauszer \cite{Rauszer74,Rauszer77} (called {\em Heyting-Brouwer logic}), 
who defined also its Kripke semantics; a category-theoretical approach to the topic 
is due to Makkai, Reyes and Zolfaghari \cite{Mak-Rey,r-zo}. The suggestion by 
F.~W.~Lawvere to use co-Heyting algebras as a logical framework to treat 
the topological notion of boundary has not been fully explored yet (but see
recent work by Pagliani \cite{Pagliani}).

\medskip

\noindent
Early research showed that the extension of first order intuitionistic logic with subtraction 
yields an {\em intermediate} logic of {\em constant domains} \cite{LopezEscobar}. 
In a rich and interesting paper \cite{Crol01} T.~Crolard showed, essentially by 
Joyal's argument, that Cartesian closed categories with exponents and co-exponents 
are degenerate; in fact even the topological models of bi-intuitionistic logic, i.e., {\em bi-topological spaces}, are degenerate. Crolard's motivations are mainly computational: he studies bi-intuitionistic 
logic in the framework of the {\em classical} $\lambda\mu$ calculus, to provide a type-theoretic 
analysis of the notion of {\em coroutine}; then he identifies a subclass of {\em safe coroutines} 
that can be typed constructively \cite{Crol04}. From our viewpoint, Crolard's work suggests two directions
of research. On one hand, it opens the way to a ``bottom up'' approach to safe coroutines, independent of the $\lambda\mu$ calculus, i.e., co-intuitionistic coroutines 
\cite{Bellin05,Bellin12,BellinLAM,BellinMenti}. On the other hand, the question arises whether the collapse of algebraic and topological models may be avoided by building the intuitionistic and co-intuitionistic sides separately, starting from distinct sets of elementary formulas, and then by joining the two sides with mixed connectives (mainly, two negations expressing the duality): this is our variant of bi-intuitionistic logic, presented in \cite{Bellin04,Bellin12,BellinCarraraChiffiMenti}.

\medskip

\noindent
Both of these tasks were advocated by this author and pursued within a project of ``logic 
for pragmatics'' with motivations from linguistics and natural language representation
\cite{Bellin12,Bellin04,BellinDPozza}. In the characterization of the logical properties of 
``illocutionary acts'', such as {\em asserting}, {\em making hypotheses} and {\em conjectures} 
one finds in natural reasoning forms of duality that can be related to intuitionistic dualities. 
For co-intuitionistic logic Crolard's term assignment has been adapted to a sequent-style natural deduction setting with single-premise and multiple-conclusions. For (our variant of) bi-intuitionistic logic Kripke semantics has been given (both in {\bf S4} and in {\em bi-modal} {\bf S4}) and a sequent calculus has been proposed where sequents are of the form 
$$
\Gamma\ ;\ \Rightarrow\ A\ ;\ \Upsilon \qquad \hbox{or}\qquad \Gamma\ ;\ C\ \Rightarrow\ ;\ \Upsilon
$$
where $\Gamma$ and $A$ are intuitionistic (assertive) formulas and $C$ and $\Upsilon$ co-intuitionistic
(hypothetical). 
 
\medskip

\noindent
But from the viewpoint of category theory a crucial remark by Crolard shows that already in 
{\em co-intuitionistic} logic there is a problem: namely, only trivial co-exponents exist in the category 
{\bf Set}. Indeed the categorical semantics of intuitionistic disjunction is given by 
{\em coproducts} \cite{LamScott}, which in {\bf Set} are represented by {\em disjoint unions}. 
On the other hand the categorical semantics of subtraction is given by {\em co-exponents}.
The co-exponent of $A$ and $B$ is an object $B_A$ together with an arrow 
$\ni_{A,B}: B \rightarrow B_A\oplus A$
such that for any arrow $f: B\rightarrow C\oplus B$ there exists a unique
$f_*: B_A \rightarrow C$ such that the following diagram commutes:
 $$
\xymatrix{B\ar[r]^f \ar[dr]_{\ni_{A,B}} & C\oplus A \\
           & B_A \oplus A \ar[u]_{f_*\oplus{id}_A}\\}
$$
It follows that 
\begin{quotation}
\noindent
{\em in the category of sets, the co-exponent $B_A$ of two sets $A$ and $B$ is defined if and only if $A = \emptyset$ or $B = \emptyset$}
 (see \cite{Crol01}, Proposition 1.15).\end{quotation}
The proof is instructive: in {\bf Set}, the coproduct $\oplus$ is {\em disjoint union}; thus if 
$A \neq \emptyset \neq B$ then the functions $f$ and $\ni_{A,B}$ for every $b\in B$ must 
{\em choose a side}, left or right, of the coproduct in their target and moreover $f_{\star}\oplus 1_A$ leaves the side unchanged. Hence, if we take a nonempty set $C$ and $f$ with the property that for some $b$ different sides are chosen by $f$ and $\ni_{A,B}$, then the diagram does not commute.   

\medskip

\noindent
Thus to have a categorical semantics of co-exponents we need categories where a different notion
of disjunction is modelled. The connective {\em par} of linear logic is a good candidate and a treatment
of {\em par} is available in {\em full intuitionistic linear logic} ({\bf FILL}) \cite{BraunerPaiva,HylandPaiva}, with a proof-theory and a categorical semantics. The {\em multiple-conclusion} consequence relation of 
{\bf FILL} and its term assignment have given motivation and inspiration to our work, as a calculus
where a distinct term is assigned to each formula in the succedent. The language of {\bf FILL} has
tensor ($\otimes$), linear implication $(\limp)$ and {\em par} $(\wp)$ and a main proof-theoretic  
concern has been the compatibility between {\em par} and \emph{linear implication}, namely, to find 
restrictions on the introduction of linear implication that guarantee its \emph{functional}  intuitionistic character and at the same time allow to prove cut-elimination (on this point see also \cite{Bellin97}). 

\medskip

\noindent
Linear co-intuitionistic logic appears already in Schellinx\cite{Schellinx} and Lambek \cite{Lambek95}. Works by R.~Blute,~J.~Cockett,~R.~Sely and T.~Trimble on \emph{wekly distributive categories} \cite{BCS,BCST} provide a sophisticated technology of natural deduction, proof-nets and categorical models for various systems of  linear logic without an involutory negation; Cockett and Seely 
\cite{CockettSeely} consider also non-commutative systems with implications and subtractions. When 
\emph{boxes} or other conditions are given for intuitionistic linear implications and co-intuitionistic subtractions, these systems provide a suitable categorical proof theory of linear bi-intuitionism. 

\medskip

\noindent
To construct categorical models of linear co-intuitionistic logic it suffices to notice that in  {\em monoidal categories} {\em par} can be modelled by a monoidal operation and co-exponents as the {\em left adjoint} of {\em par}. The main task then is to model Girard's exponential {\em why not?}: in this way a categorical semantics for co-intuitionistic logic can be recovered by applying the dual of Girard's translation of intuitionistic logic into linear logic, namely: 
\begin{center}
\begin{tabular}{rl}
$(p)^{\circ}$ =& p\\
$(\mathbf{f})^{\circ}$ = &{\bf 0}\\
$(C \curlyvee D)^{\circ}$ =& $?(C^{\circ}\oplus D^{\circ})$ = $?(C^{\circ}) \wp ?(D^{\circ})$\\
$(C \smallsetminus D)^{\circ}$ = & $C^{\circ}\smallsetminus (? D^{\circ})$\\
$(E\vdash C_1, \ldots, C_n)^{\circ}$ =& $?(E^{\circ})\vdash ?(C_1^{\circ}), \ldots, ?(C_n^{\circ}))$\\
\end{tabular}
\end{center}
where {\bf 0} is the identity of $\oplus$ and we use ``$\smallsetminus$'' both in linear and in
non-linear co-intuitionistic logic.

\medskip

\noindent
The task amounts to dualizing Nick Benton, Gavin Bierman, Valeria de Paiva and Martin Hyland's
well-known semantics for intuitionistic linear logic \cite{BBHdP92}. This may be regarded as a routine
exercise, except that one has to provide a term assignment suitable for the purpose. In this task we build on a term assignment to multiplicative co-intuitionistic logic, which has been proposed as an abstract {\em distributed calculus} dualizing the linear $\lambda$ calculus 
\cite{BellinLAM,Bellin05,BellinMenti}: in our view such a dualization underlies the translation of 
the linear $\lambda$-calculus into the $\pi$-calculus (see \cite{BellinScott}). 

\medskip

\noindent 
As a matter of fact, Nick Benton's {\em mixed Linear and Non-Linear} logic \cite{Benton95}
may give us not only an easier approach to modelling the exponentials but also the key to 
a categorical semantics of (our version of) bi-intuitionistic logic: indeed, by dualizing the linear part 
of Benton's system we may obtain both a proof-theoretic and a category theoretic framework for 
mixed co-intuitionistic linear and intuitionistic logic and thus also for bi-intuitionistic logic - of course, 
we need to use the exponential {\em why not?} and dualize Girard's translation. But then a categorical investigation of {\em linear cointuitionistic logic} and of the {\em why not?} exponential is a preliminary step in this direction and has an independent interest.
\section{Proof Theory}
The language of {\em co-intuitionistic linear logic}, given an infinite sequence of elementary formulas 
$\eta_1, \eta_2, \ldots$, is defined by the following grammar:\footnote{In accordance with our 
interpretation of co-intuitionism as a logic of hypotheses, we may write elementary formulas $\eta$ 
as $\hyp p$, where ``$\hyp\;$'' is a sign for the illocutionary force of hypothesis and $p$ is an atomic 
proposition. Such a linguistic analysis plays no explicit role in this paper.}
\[
C, D \ :=\ \eta\ |\ \bot\ |\ C \wp D\ |\ C \smallsetminus D\ |\ ? C
\]
The rules of sequent-style natural deduction {\bf co-ILL} for co-Intuitionistic Linear Logic 
are given in Table \ref{cointnatded}.
\begin{table}[ht]
 \begin{center}
  {\footnotesize
   \begin{tabular}{|cc|}
   \hline
&\\
{\it assumption}&{\it substitution}\\
$A\vdash A$ &
\AxiomC{$E\vdash \Gamma, A$}
\AxiomC{$A\vdash \Delta$}
\BinaryInfC{$E\vdash\Gamma, \Delta$}
\DisplayProof
\\
&\\
$\bot$-I &$\bot$-E\\
\AxiomC{$E \vdash \Gamma$}
\UnaryInfC{$E \vdash \Gamma, \bot$}
\DisplayProof
&
$\bot\vdash$\\ 
& \\
$\smallsetminus$-I &$\smallsetminus$-E\\
\AxiomC{$E\vdash \Gamma, C$}
\AxiomC{$D\vdash \Delta$}
\BinaryInfC{$E\vdash \Gamma, C\smallsetminus D, \Delta$}
\DisplayProof&
\AxiomC{$H \vdash \Upsilon, C\smallsetminus D$}
\AxiomC{$C \vdash D, \Delta$}
\BinaryInfC{$H \vdash \Upsilon, \Delta$} 
\DisplayProof\\
&\\
$\wp$-I&$\wp$-E\\
\AxiomC{$E\vdash \Gamma, C_0, C_1$}
\UnaryInfC{$E\vdash \Gamma, C_0\,\wp\, C_1$}
\DisplayProof\quad
&
\AxiomC{$H \vdash \Upsilon, C_0\,\wp\, C_1$}
\AxiomC{$C_0 \vdash \Gamma_0$}
\AxiomC{$C_1 \vdash \Gamma_1$}
\TrinaryInfC{$H \vdash \Upsilon, \Gamma_0,\Gamma_1$}
\DisplayProof\\
&\\
{\em dereliction} & {\em storage} \\
\AxiomC{$E\vdash \Gamma,\, C$}
\UnaryInfC{$E\vdash \Gamma,\, ?C$}
\DisplayProof
&
\AxiomC{$H\vdash \Upsilon,\, ?C$}
\AxiomC{$C\vdash ?\Delta$}
\BinaryInfC{$H\vdash \Upsilon,\, ?\Delta$}
\DisplayProof\\
& \\
{\em weakening} & {\em contraction} \\
\AxiomC{$E\vdash \Gamma$}
\UnaryInfC{$E\vdash \Gamma,\, ?C$}
\DisplayProof
&
\AxiomC{$E\vdash \Gamma,\, ?C, ?C$}
\UnaryInfC{$E\vdash \Gamma,\, ?C$}
\DisplayProof\\
& \\
\hline
  \end{tabular}}
  \caption{Natural Deduction for {\bf co-ILL}}
  \label{cointnatded}
 \end{center} 
\end{table}
As {\em co-intuitionistic linear logic} may be quite unfamiliar, we sketch an intuitive explanation 
of its proof theory. We think of co-intuitionistic logic as being about {\em making hypotheses}
\cite{Bellin12,Bellin04,BellinCarraraChiffiMenti}. It has a consequence relation of the form 
\begin{equation} \label{Example:0}
H \vdash H_1, \ldots, H_n.
\end{equation} 
Suppose $H$ is a hypothesis: which (disjunctive sequence of) hypotheses {\em $H_1$ or 
$\ldots$ or $H_n$} follow from $H$?  Since the logic is {\em linear}, 
commas in the meta-theory stand for Girard's {\em par} and the structural rules Weakening 
and Contraction are not allowed. A relevant feature, which we shall not discuss here, 
is that the consequence relation may be seen as {\em distributed}, i.e., we may think of 
the alternatives $H_1$, $\ldots$, $H_n$ in (\ref{Example:0}) as lying in different {\em locations}
\cite{BellinLAM,BellinMenti}. 

\medskip

\noindent
The main connectives are {\em subtraction} $A\smallsetminus B$ ({\em possibly $A$ and 
not $B$}) and Girard's {\em par} $A\wp B$. Natural Deduction inference rules for {\em subtraction}
(in a sequent form) are as follows.
 $$
\AxiomC{$H\vdash \Gamma, C$}
\AxiomC{$D\vdash \Delta$}
\LeftLabel{$\smallsetminus$-{\em intro}}
\BinaryInfC{$H\vdash \Gamma, C\smallsetminus D, \Delta$}
\DisplayProof
\qquad
\AxiomC{$H\vdash \Delta, C\smallsetminus D$}
\AxiomC{$C\vdash D, \Upsilon$}
\LeftLabel{$\smallsetminus$-{\em elim}}
\BinaryInfC{$H\vdash \Delta, \Upsilon$}
\DisplayProof
$$
Notice that in the $\smallsetminus$-{\em elimination} rule the evidence that 
$D$ may be derivable from $C$ given by the {\em right premise} has become 
{\em inconsistent with the hypothesis} $C\smallsetminus D$ in the left premise;
in the conclusion we drop $D$ and we {\em set aside} the evidence for the  
inconsistent alternative. Namely, such evidence is not destroyed, but rather 
stored somewhere for future use. 

\medskip

\noindent
If the {\em left premise} of $\smallsetminus$-{\em elimination}, deriving  
$C\smallsetminus D$ or $\Delta$ from $H$, has been obtained 
by a $\smallsetminus$-{\em introduction}, this inference has the form 
$$\displaystyle{\AxiomC{$H\vdash \Delta_1, C$}
\AxiomC{$D\vdash \Delta_2$}
\BinaryInfC{$H\vdash\Delta_1, \Delta_2, C\smallsetminus D$}
\DisplayProof}.$$ Then the pair of {\em introduction/elimination} rules can be eliminated: 
using the {\em removed evidence} that $D$ with $\Upsilon$ are derivable from $C$ 
({\em right premise} of the $\smallsetminus$-{\em elim.})
we can conclude that $\Delta_1, \Delta_2, \Upsilon$ are derivable from $H$. 
This is, in a nutshell, the principle of normalization (or {\em cut-elimination}) for subtraction. 

\medskip

\noindent
The {\em storage} operation is made explicit in the rules for the {\bf ?} operator of linear logic.
Here an entire derivation $d$ of $?\Delta$ from $C$ (where $? \Delta = ?D_1,\ldots, ?D_n$)  
is set aside; what is accessible now is something like a non-logical axiom of the form 
$?C \vdash ?\Delta$. However in the process of normalization the derivation $d$ may be
recovered to be {\em used}, {\em discarded} or {\em copied} in the interaction of a {\em storage}
rule with {\em dereliction}, {\em weakening} or {\em contraction}: all of this is conceptually clear, 
thanks to J-Y.~Girard, and has been mathematically analyzed in the {\em geometry of interaction.}
\subsection{From Crolard's classical coroutines to co-intuitionistic ones}
Crolard \cite{Crol04} provides a term assignment to the subtraction rules in 
the framework of Parigot's $\lambda\mu$-calculus, typed in a sequent-style 
natural deduction system. The $\lambda\mu$-calculus provides a typing system 
for functional programs with {\em continuations} and a computational interpretation 
of classical logic  (see, e.g., \cite{Curien,Selinger}). 

\medskip

\noindent
In the type system for the $\lambda\mu$ calculus sequents may be written 
in the form $\Gamma \vdash t: A\ |\ \Delta$, with contexts $\Gamma$ = 
$x_1:C_1, \ldots, x_m:C_m$ and $\Delta$ = $\alpha_1:D_1, \ldots, \alpha_n:D_n$,
where the $x_i$ are {\it variables} and the $\alpha_j$ are $\mu$-{\it variables}
(or {\em co-names}). 
In addition to the rules of the simply typed lambda calculus, there are 
{\it naming rules} 
$$
{{\Gamma \vdash t: A\ |\ \alpha: A, \Delta}\over%
{\Gamma \vdash [\alpha]t: \bot\ |\ \alpha: A, \Delta}}[\alpha]%
\hskip1in %
{{\Gamma \vdash t: \bot\ |\ \alpha: A, \Delta}\over{\Gamma \vdash %
\mu\alpha.t: A\ |\ \Delta}}\mu
$$
whose effect is to ``change the goal'' of a derivation and which allow us 
to represent the familiar {\em double negation rule} in Prawitz Natural Deduction. 

\medskip

Crolard extends the $\lambda\mu$ calculus with introduction and elimination 
rules for subtraction:\footnote{Actually in Crolard \cite{Crol04} the introduction rule is given 
in the more general form of $\smallsetminus$-{\em introduction} with two sequent premises 
(which we use below) and more general {\it continuation contexts} occur in place of $\beta$; 
the above formulation is logically equivalent and suffices for our purpose.}
{\small
\begin{center}
\begin{tabular}{c}
$\displaystyle{\strut{\Gamma \vdash t: A\ |\ \Delta}\over{\strut%
\Gamma \vdash \hbox{\tt make-coroutine}(t, \beta) : A \smallsetminus B\ |\ %
\beta: B, \Delta}}%
\smallsetminus I\hskip1in$\\
\\
$\hskip1in\displaystyle{\strut{\Gamma \vdash t: A\smallsetminus B\ |\ \Delta %
 \qquad \Gamma,\ x:A\vdash u: B\ |\ \Delta}%
\over{\strut%
\Gamma \vdash \mathtt{resume}\ t\ \mathtt{with}\ x\mapsto u: C\ |\ \Delta}}%
\smallsetminus E$\\
\end{tabular}
\end{center}}

\noindent
The reduction of a redex of the form 
\begin{center}
{\small
\AxiomC{$\Gamma\vdash t: A\ |\ \Delta$}
\RightLabel{$\smallsetminus$-I}
\UnaryInfC{$\Gamma \vdash %
\hbox{\tt make-coroutine}(t,\beta): A\smallsetminus B\ |\ %
\beta: B, \Delta$}
\AxiomC{$\Gamma, x: A \vdash u: B\ |\ \Delta$}
\RightLabel{$\smallsetminus$-E}
\BinaryInfC{$\Gamma \vdash %
\mathtt{resume}\ (\hbox{\tt make-coroutine}(t,\beta) )\, %
\mathtt{with}\ x\mapsto u\ : C\ |\ \beta: B, \Delta$}
\DisplayProof
}
\end{center}
is as follows: 
\begin{center}
{\small
\AxiomC{$\Gamma\vdash t: A\ |\ \Delta$}
\AxiomC{$\Gamma, x: A \vdash u\; B\ |\ \Delta$}
\RightLabel{{\it substitution}}
\BinaryInfC{$\Gamma \vdash u[t/x]: B\ |\ \Delta$}
\RightLabel{$[\beta]$}
\UnaryInfC{$\Gamma \vdash %
[\beta] u[t/x]: \bot\ |\ \beta: B, \gamma: C, \Delta'$}
\RightLabel{$\mu$}
\UnaryInfC{$\Gamma \vdash \mu\gamma.[\beta] u[t/x]: C\ |\ \beta: B, \Delta'$}
\DisplayProof
}
\end{center}
Notice that Crolard's elimination rule involves an application of the rule 
\emph{ex falso quodlibet}, which is explicit in the definition of the operator 
$\mathtt{resume}$ (see \cite{Crol04}, Remark 4.3.). This seems unavoidable 
working within the $\lambda\mu$ calculus, but it may not be desirable 
outside that framework.

\medskip

\noindent
Working with the full power of classical logic, if a constructive system of 
bi-intuitionistic logic is required, then the implication right and 
subtraction left rules must be restricted; this can be done by considering 
{\it relevant dependencies}.\footnote{For instance, in the derivation of the 
right premise $\Gamma, x:A \vdash u:B\ |\ \Delta$
of a subtraction elimination ($\smallsetminus E$), there should be no relevant 
dependency between the formula $B$ and the assumptions in $\Gamma$, but only 
between $B$ and $A$. Similar issues arise in {\bf FILL}, see \cite{HylandPaiva} and
\cite{Bellin97}, section 4.} 
Crolard is able to show that the term assignment for such a restricted logic is a calculus of 
{\it safe coroutines}, described as terms 
in which no coroutine can access the local environment of another coroutine. 

\medskip

\noindent
Crolard's work suggests the possibility of defining co-intuitionistic coroutines directly,
independently of the typing system of the $\lambda\mu$-calculus. Since $\mu$-variable 
abstraction and the $\mu$-rule are devices to change the ``actual thread'' of computation, 
the effect of removing such rules is that all ``threads'' of computation are simultaneously
represented in a multiple conclusion sequent, but variables $y$ that are temporarily 
inaccessible in a term $N$ are being replaced by a term $\mathtt{y}(M)$ by the substitution 
$N[y:= \mathtt{y}(M)]$, where $M$ contains a free variable $x$ which is accessible in the 
current context. This is the approach pursued in \cite{Bellin12,Bellin05,Bellin04} leading to 
the present categorical presentation.
\subsection{A dual linear calculus for $\mathbf{MNJ}^{\smallsetminus\wp\bot}$}
\label{ConjecturalTermAssign}
We present the grammar and the basic definitions of our dual linear calculus for linear co-intuitionistic logic with subtraction, disjunction and \emph{why not?} ({\bf ?}) operator. 
\begin{defi}\label{lineartermdef}
We are given a countable set of {\em free variables} (denoted by $x$, $y$, $z$ $\ldots$), 
and a countable set of {\em unary functions} (denoted by $\mathtt{x}, \mathtt{y}, \mathtt{z}, \ldots$).
The terms of our calculus, denoted by $R$, are either \emph{$m$-terms}, denoted by $M$, $N$, 
or \emph{$p$-terms}, denoted by $P$. 

\medskip

\noindent
(i) \emph{Multiplicative} terms, $m$-terms and $p$-terms are defined by the following grammar.%
\begin{center}
\begin{tabular}{rl}
$M, N\ :=$& %
$ x\ |\ \mathtt{x}(M) \ |\ \mathtt{connect\ to}(R)\ |\ M\wp N\ |\ \mathtt{casel}(M)\ |\ \mathtt{caser}(M)\ |$\\
& $|\ \hbox{\tt make-coroutine}(M, \mathtt{x})\ |\ [] . $ \\ 
$P\ :=$& $\mathtt{postpone}(\mathtt{y}\mapsto N, M)\ |\ \mathtt{postpone}(M) .$\\ 
 $R\ :=$& $M\ |\ P.$\\  
\end{tabular}
\end{center}
\medskip

\noindent
(ii)\emph{Multiplicative and exponential} terms, $m$-terms and $p$-terms are obtained by adding to 
the above grammar the following clauses: 
\begin{center}
\begin{tabular}{rl}
$M, N\ :=$& $\ldots \ |\ [M]\ |\ [M,N]$.\\
$P\ :=$& $\ldots\ |\  \mathtt{store}(P_1, \ldots, P_m, M_1, \ldots, M_n, \mathtt{y}_1, \ldots, \mathtt{y}_n, \mathtt{x}, N)$\\
\end{tabular}
\end{center}
\end{defi}
We usually abbreviate ``$\mathtt{make-coroutine}$'' as ``$\mathtt{mkc}$'' and ``$\mathtt{postpone}$''
as ``$\mathtt{postp}$''. We often write $\overline{P}$ for a list  $P_1, \ldots, P_m$, and similarly 
$\overline{M}$ for $M_1, \ldots, M_n$.  ``$[]$'' is the empty term ($\mathtt{nil}$). 
\begin{defi} The {\em free variables} $FV(M)$ in a term $R$ are defined thus:  
\begin{small}
\begin{center}
\begin{tabular}{rl} 
$FV(x)$\ =\ &$\{x\}$\\
$FV(\mathtt{x}(M))$\ =\ &$FV(M)$\\
$FV(\mathtt{connect\ to}(R))$\ =\ &$FV(R)$\\
$FV(M\wp N)$\ =\ & $FV(M) \cup FV(N)$\\
$FV(\mathtt{casel}(M)) = FV(\mathtt{caser}(M))$\ =\ & $FV(M)$\\
$FV(\hbox{\tt mkc}\, (M, \mathtt{x}))$\ =\ & $FV(M)$\\
$FV(\mathtt{store}(\overline{P}, M_1, \ldots, M_n, \mathtt{y}_1,\ldots, \mathtt{y}_n, \mathtt{x}, N))$\ =\ &
$(FV(\overline{P})\cup (FV(\overline{M}) ) \smallsetminus \{x\}) \cup FV(N)$\\
& where $\overline{M} = M_1, \ldots, M_n$.\\
$FV(\,[]\,) = \emptyset\quad FV(\,[M]\,) = FV(M)\quad$ &  $FV(\,[\,M,N\,]\,)\ =\ FV(M)\cup FV(N)$\\
$FV(\hbox{\tt postp}(\mathtt{x}\mapsto N, M))$\ =\ 
&$(FV(N)\smallsetminus \{x\})\cup FV(M)$.\\
$FV(\hbox{\tt postp}(M))$\ =\ & $FV(M)$. 
\end{tabular}
\end{center}
\end{small}
\end{defi}
\begin{defi} Let $\|$ be a binary operation on terms (\emph{parallel composition})  which is associative, commutative and has the empty term $[]$ as the identity. Terms generated by (zero or more) 
applications of parallel composition are called \emph{contexts}. Thus \emph{contexts} are generated by the following grammar: 
\begin{center}
\begin{tabular}{rl}
$C\ :=$& $R\  |\ (C \| R)$\\ 
\end{tabular}
\end{center}
\emph{modulo the structural congruences}
\begin{enumerate}[label=(\roman*)]
\item \qquad $R_0 \| (R_1 \| R_2)\equiv  (R_0 \| R_1) \| R_2$,
\item \qquad $R_0 \| R_1 \equiv  R_1 \| R_0$,
\item \qquad $(R_0 \| []) \equiv  R_0$, 
\item \qquad $C_0 \| R \| C_1 \equiv C_0 \| R'\| C_1$ if $R \equiv R'.$
\end{enumerate}
\end{defi}

\noindent
Let $R_1\| \ldots \| R_k$ be a context, where all $R_i$ are non-null , $i\leq k$. 
Notice that the notation is well-defined by generalized associativity. 
We write $\mathcal{S}_{\overline{x}} : R_1\| \ldots \| R_k$ if all free variables occurring in 
$R_1, \ldots, R_k$ are in the list $\overline{x}$.

\medskip

\noindent
\emph{Computational contexts}, the basic expressions of our calculus, are contexts satisfying 
some correctness conditions, that guarantee the identification of a context and rule out circular 
structures.  Our ``calculus of coroutines'' is used here in a typed setting, where self referential 
structures are not needed. 
\begin{defi}\label{acyclicity} 
An expression $\mathcal{S}_x : R_1\| \ldots \| R_k$ is a (correct) \emph{computational context} 
if it satisfies the following axioms. 
\begin{enumerate}
\item\label{ac:one} Each term in the set $\{R_1, \ldots, R_k\}$ contains $x$ and no other free variable.
\item\label{ac:two} In every term of the form $\hbox{\tt postp}(\mathtt{y}\mapsto N, M)$ the term $N$ 
contains a free variable $y$ with $y \notin FV(M)$ and no other free variable. 
\item\label{ac:three} In every term 
$\mathtt{store}(\overline{P}, N_1, \ldots, N_n, \mathtt{y}_1, \ldots, \mathtt{y}_n, \mathtt{z}, M)$
the terms $N_i$ are of the forms $[N]$ or $\mathtt{connect\ to}(R)$ for some $N$ or $R$. 
\item\label{ac:four} Let $\mathbf{S}=  %
\mathtt{store}(\overline{P}, N_1, \ldots, N_n, \mathtt{y}_1, \ldots, \mathtt{y}_n, \mathtt{z}, M)$
occur within a multiplicative computational context $\mathcal{S}_x$; write $\mathcal{S}^-_x$ for
 $\mathcal{S}_x$ without $\mathbf{S}$. Then $\{\overline{P}, N_1, \ldots, N_n\}$ is a computational context $\mathcal{S}_z$ for some free variable $z$ with $z \neq x$. 
We say that $\mathcal{S}_z$ \emph{occurs immediately within} $\mathbf{S}$. 
\item\label{ac:five} In a computational context $\mathcal{S}_x$ the nesting of $p$-terms of the form 
$\mathtt{store}$ within $\mathcal{S}_x$ has the structure of a rooted tree, with root 
$\mathcal{S}^-_x$ itself.
\end{enumerate}
A computational context is said \emph{multiplicative} if it does not contain $\mathtt{store}$ terms.  
\end{defi}
\begin{rem} By axiom \ref{ac:one} the relevant components of a computational context are uniquely identified. Axiom \ref{ac:two} is analogue to the \emph{acyclicity} condition in proof nets for linear logic. Axioms \ref{ac:four} and \ref{ac:five} induce a structure on context that corresponds to that of 
\emph{boxes} in proof nets. Axiom \ref{ac:three} characterizes \emph{exponential boxes} in our framework. 
\end{rem}
\begin{defi}[$\alpha$-equivalence]\label{alpha}
Let $\mathcal{S}_x$ and $\mathcal{S}_{x'}$ be computational contexts.
To define what it means that  $\mathcal{S}_x$ and $\mathcal{S}_{x'}$ are $\alpha$-equivalent, 
we need to define this property for sets of terms $\mathcal{S}_{\overline{x}}$ and 
$\mathcal{S}_{\overline{x}'}$ that may contain more than one free variable from the lists $\overline{x}$
and $\overline{x}'$, respectively; therefore they are not correct computational contexts. 
The definition is by induction on the number of terms occurring in $\mathcal{S}_{\overline{x}}$. 
\begin{enumerate}
\item If $\mathcal{S}_x = \{x\}$, then $\mathcal{S}_x
\equiv\mathcal{S}_{x'}$ iff $\mathcal{S}_{x'} = \{x'\}$ and $x = x'$;
\item If $\mathcal{S}_{\overline{x}} = \{\mathtt{x}(M)\}$, then 
$\mathcal{S}_{\overline{x}} \equiv\mathcal{S}_{\overline{x}'}$ iff
$\mathcal{S}_{\overline{x}'} =   \{\mathtt{x}(M')\}$ and $M \equiv M'$. A similar definition applies 
if $\mathcal{S}_{x} =  \{\mathtt{connect\ to}(M)\}$ or 
$\{\mathtt{casel}(M)\}$ or $\{\mathtt{caser}(M)\}$ or $\{ [M] \}$ or $\{\mathtt{postp}(P)\}$; 
\item  If $\mathcal{S}_{\overline{x}}  = \{ M\wp N\}$, then 
$\mathcal{S}_{\overline{x}} \equiv\mathcal{S}_{\overline{x}'}$ iff 
$\mathcal{S}_{\overline{x}'}  = \{ M'\wp N'\}$  and $M \equiv M'$ and $N \equiv N'$. A similar definition 
applies if $\mathcal{S}_x  = \{[M, N] \}$. 
\item Let $\mathcal{S}_{\overline{x}}$ be partitioned as 
\[
\mathcal{S}^-_{\overline{x}} \cup\; \{\mathtt{mkc}(M, \mathtt{y})\}\;
\cup\; \mathcal{S}_{\overline{x}y} [y:= \mathtt{y}(M)];
\] 
then $\mathcal{S}_{\overline{x}} \equiv\mathcal{S}_{\overline{x'}}$ iff 
$\mathcal{S}_{\overline{x}'}$ can be partitioned as $\mathcal{S}^-_{\overline{x}'} \cup
\{\mathtt{mkc}(M', \mathtt{y})\} \cup \mathcal{S}_{\overline{x}'y'} [y':= \mathtt{y}(M')]$
and $\mathcal{S}^-_{\overline{x}} \cup \{M\} \equiv \mathcal{S}^-_{\overline{x}'} \cup \{ M'\}$ and, moreover, for all variables $v$ except for a finite number 
$\mathcal{S}_{\overline{x}y} [y := v] \equiv\mathcal{S}_{\overline{x}'y'} [y' := v ]$. 
\item Let $\mathcal{S}_{\overline{x}}$ can be partitioned as 
\[
\mathcal{S}^-_{\overline{x}}\cup \{\mathtt{postpone}(\mathtt{y}\mapsto N, M)\}\cup 
\mathcal{S}_{\overline{x}y} [y:= \mathtt{y}(M)];
\] 
then $\mathcal{S}_{\overline{x}} \equiv\mathcal{S}_{\overline{x}'}$ iff  $\mathcal{S}_{\overline{x}'}$ 
can be partitioned as 
$\mathcal{S}^-_{\overline{x'}} \cup (\{\mathtt{postpone}(y'\mapsto N', M')\}\cup %
\mathcal{S}_{\overline{x}'y'}) [y':= \mathtt{y}'(x')]$ and 
$\mathcal{S}^-_{\overline{x}} \equiv \mathcal{S}^-_{\overline{x}'}$ and, moreover, for all variables $v$ except for a finite number 
$(\mathcal{S}_{\overline{x}y}\cup \{N\} )[y:= v] \equiv (\mathcal{S}_{\overline{x}'y'}\cup \{N'\} )[y':= v]$.
\item Let $\mathcal{S}_{\overline{x}}$ be partitioned as 
$\mathcal{S}^-_{\overline{x}}\cup \{\mathbf{S}_1, \ldots, \mathbf{S}_k \}$ 
where for $i \leq k$, $\mathbf{S}_i$ is a $\mathtt{store}$ term with a set of terms   
$\mathcal{S}_{\overline{x}z_i}$ immediately inside it. 
 Then $\mathcal{S}_{\overline{x}} \equiv\mathcal{S}_{\overline{x}'}$ 
iff $\mathcal{S}_{\overline{x}'}$ can be partitioned in a similar way as 
$\mathcal{S}^-_{\overline{x}'} \cup \{\mathbf{S}'_1, \ldots, \mathbf{S}'_k \}$ where for $i \leq k$ the set  $\mathcal{S}_{\overline{x'}z'_i}$ occurs immediately inside $\mathbf{S'}_i$, for $i \leq k$ and, moreover
\begin{enumerate}[label=(\roman*)]
\item
$\mathcal{S}^-_{\overline{x}}\equiv \mathcal{S}^-_{\overline{x}'}$ 
\item for all $i \leq k$ and for all variables $v$ except for a finite number 
$\mathcal{S}_{\overline{x}z_i}[z_i := v] \equiv \mathcal{S}_{\overline{x}'z'_i}[z'_i := v]$.
\end{enumerate}
\end{enumerate}
\end{defi}
\begin{rems}\label{local-remote-binding} 
(i) Consider the terms $\mathtt{make}-\mathtt{coroutine}$, binary $\mathtt{postpone}$ and $\mathtt{store}$. 
They are binders acting on a whole computational context, rather than on a delimited scope 
within a single term. 
We may call their action \emph{remote binding}; it is expressed by a substitution of some $m$-term 
$\mathtt{x}(M)$ for the free variable $x$ throughout a computational context $\mathcal{S}_x$. One could express remote binding by a more familiar notation, as in the $\lambda$-calculus or in the $\pi$-calculus; but then the scope of a binder would partition a context and parallel composition would not appear as a 
top-level operator only. In the typed case one could not assign a separate $m$-term to each formula in the succedent: at best, one could assign an "access port'' to a unique term assigned to the whole sequent, as in the translations of linear logic into the $\pi$-calculus (see \cite{BellinScott}). This would be
against a main motivation of this calculus, namely, to give a ``distributed term assignment'' for a ``multiple conclusion'' co-intuitionistic deductive system. An application of our notation for ``remote binding'' is found in section \ref{ProbabilisticInterpretation} on a probabilistic interpretation of \emph{subtraction} and \emph{par}. 

\medskip

\noindent
(ii) The $p$-terms binary $\mathtt{postpone}$ and $\mathtt{store}$ are also \emph{local binders} of the free variable occurring in their argument. Indeed in a term $\mathtt{postpone}(\mathtt{y}\mapsto N, M)$ a free variable $y$ occurring in $N$ becomes locally bound; then $y$ is replaced by $\mathtt{y}(M)$ in the computational context, to express remote binding. In a term 
$\mathtt{store}(\overline{P}, N_1, \ldots, N_n, \mathtt{y}_1, \ldots, \mathtt{y}_n, \mathtt{z}, M)$
a free variable $z$ becomes bound; then a term $\mathtt{z}(N_i)$ occurs in the term 
$\mathtt{y}_i(\mathtt{z}(N_i))$, that replaces the stored $N_i$ in the computational context. 
A more complete notation of our p-terms  would be 
$\mathtt{postpone}(y\mapsto N, M)\;\mathtt{with\ y\ for}\ y$ and 
$\mathtt{store}(\overline{P}, N_1, \ldots, N_n, \mathtt{y}_1, \ldots, \mathtt{y}_n, M)\ %
\mathtt{with\ z\ for}\ z$, explicitly establishing the connection between the locally bound variable and 
its corresponding unary function. For terms of the form $\mathtt{store}$ we are spared the more verbose notation by the \emph{acyclicity axioms}, by which the locally bound variable is uniquely identified. 

\medskip

\noindent
(iii) In the notation of $p$-terms of the form $\mathtt{postpone}(y\mapsto N, M)$ it is sometimes convenient to ignore the distinction between \emph{local binding} and \emph{remote binding} and treat the occurrences in $N$ of the variable $y$ as remotely bound; namely, we can write  
$\mathtt{postpone}(\mathtt{y}\mapsto N [y := \mathtt{y}(M)], M)$. Indeed the distinction between local and remote binding can be recovered from the structure of the terms. This notation allows us to adopt a 
tree-like notation for decorated co-intuitionistic Natural Deduction derivations analogue to that of Prawitz's Natural Deduction trees decorated with $\lambda$-terms. On the other hand, such a treatment seems unmanageable for p-terms of the form $\mathtt{store}$; for them we retain a \emph{box-like} notation. 
\end{rems}
\begin{defi}
Substitution of a term $t$ for a free variable $x$ in a term $R$ is defined as follows: 
\begin{center}
\begin{tabular}{rl}
$x[x:= M] = M,\quad$&\ $y[x:= M] = y$ if $x\neq y$;\\
$\mathtt{connect\ to}(R)[x:= M]\ $=&\ $\mathtt{connect\ to}(R[x:= M])$\\ 
$\mathtt{postp}(N)[x:= M]\ $=&\ $\mathtt{postp}(N[x:= M])$\\
$\mathtt{y}(N)[x:= M]$ =&\ $\mathtt{y}(N[x:= M])$;\\
$(N_0\wp N_1)[x:= M]$ =&\  $(N_0[x:= M]\,)\wp(\, N_1[x:= M])$\\
$\mathtt{casel}(N)[x:= M]$ =&\ $\mathtt{casel}(N[x:= M]),$\\
$\mathtt{caser}(N)[x:= M]$ =&\ $\mathtt{caser}(N[x:= M]);$\\ 
$\mathtt{mkc}(N, \mathtt{y})[x:= M]$ = &\ $\mathtt{mkc}(N[x:= M], \mathtt{y}),$\\
$\mathtt{store}(\overline{N}, \overline{\mathtt{y}}, \mathtt{z}, N)[x:= M]$ = &\
$\mathtt{store}(\overline{N}[x:= M], \overline{\mathtt{y}}, \mathtt{z}, N[x:= M)$\\ 
$\mathtt{postp}(\mathtt{y}\mapsto(N_1), N_0)[x:= M]$ = &\ 
$\mathtt{postp}(\mathtt{y}\mapsto(N_1[x:= M]), N_0[x:= M]).$\\
$[\,R\,] [x:= M] = \bigl[R[x:=M]\bigr]\quad$ &\ $[R_0, R_1][x:= M] = \bigl[R_0[x:= M], R_1[x:= M]\bigr]$\\
\end{tabular} 
\end{center}
\end{defi}
\begin{prop} \hfill

\begin{enumerate}[label=(\roman*)]
\item $\mathcal{S}_x = x$ and $\mathcal{S}_y = \mathtt{postp}(y)$ are computational contexts. 

\medskip

\item 
Let $\mathcal{S}_x = R_1\| \ldots\| R_m\| M$ and $\mathcal{S}_y = R_{m+1}\|\ldots\| R_{m+n}$ be computational contexts where $x \neq y$. We write $\mathcal{S}_y[y:= N]$ for 
$R_{m+1}[y:= N]\| \ldots\| R_{m+n}[y:= N]$. Then 
\[\mathcal{S}'_x = R_1\| \ldots\| R_m\| (\mathcal{S}_y [y:= M])\] is a computational context 
\emph{(substitution)}; 

\medskip

\item Let $\mathcal{S}_x$ and $\mathcal{S}_y$ be as in \emph{(ii)}.  Then
  $\mathcal{S}'_x = R_1\| \ldots\| R_m\| \mathtt{mkc}(M, \mathtt{y})\|
  \mathcal{S}_y [y:= \mathtt{y}(M)]$ is a computational context
  \emph{(make\ coroutine)};

\medskip

\item Let $\mathcal{S}_x = R_1\| \ldots\| R_m$ and $\mathcal{S}_y = R_{m+1}\|\ldots\| R_{m+n}$ be computational contexts. 
Then 
\[
\mathcal{S}_z = \mathcal{S}_x[x:= \mathtt{casel}(z)] \| \mathcal{S}_y[y:= \mathtt{caser}(z)]
\] 
is a computational context \emph{(cases)}; 

\medskip

\item Let $\mathcal{S}_x = R_1\| \ldots\| R_m\| M$ be a computational context and let $x \neq y$. 
Then 
\[
\mathcal{S}_y = \mathtt{postp}(x\mapsto M, y)\| R_1\| \ldots\| R_m
\] 
is a computational context \emph{(postpone)}; 

\medskip

\item Let $\mathcal{S}_x = R_1\| \ldots\| M_i\| M_{i+1}\|\ldots\| R_m$ be a computational context. Then 
\[\mathcal{S}'_x = R_1\| \ldots\| (M_i \wp M_{i+1})\| \ldots\| R_m\}
  \quad\mbox{and}\quad
  \mathcal{S}'_x = R_1\| \ldots\| [M_i,  M_{i+1}]\| \ldots\| R_m\}
\]
  are  computational contexts, 
  \emph{(par)} and \emph{(contraction)};

\medskip

\item Let $\mathcal{S}_x = R_1\|\ldots\| R_m$ be a computational context. Then 
\[\mathcal{S}'_x = R_1\| \ldots\| \ldots\| R_m\}\|
\mathtt{connect\ to}(R)\]
 is a  computational context 
\emph{(unit)} and \emph{(weakening)}; 

\medskip

\item Let $\mathcal{S}_z  = P_1 \| \ldots \| P_m\| N_1 \| \ldots \| N_n$ be a computational context 
where all terms $N_i$ are of the form $[N]$ or $\mathtt{connect\ to}(R)$ for some $N$ or $R$. 
Then 
\[\mathcal{S}_x  = 
\mathtt{store}(P_1, \ldots, P_m, N_1, \ldots, N_n, \mathtt{y}_1, \ldots, \mathtt{y}_n, \mathtt{z}, x)\]
is a computational context \emph{(store)}.
\end{enumerate}
\end{prop}
\proof In all cases the proposition is easily proved by checking that the resulting set of terms satisfies 
all the axioms in definition \ref{acyclicity}. \qed
\medskip

\noindent 
The operation of $\beta$-reduction transforms a computational context $\mathcal{S}_x$ into a
computational context $\mathcal{S}'_x$. It may be either \emph{local}, affecting only the terms where 
the \emph{redex} occurs, or a \emph{global} operation with side-effects on parts of $\mathcal{S}_x$, 
mainly relabelling the terms that express binding by $\mathtt{make-coroutine}$, $\mathtt{postpone}$ 
or $\mathtt{store}$.
\begin{defi} 
$\beta$-reduction of a {\em redex} $\mathcal{R}ed$ in a 
{\em computational context} $\mathcal{S}_x$ is defined as follows.
\begin{enumerate}[label=(\roman*)]
\item If  $\mathcal{R}ed$ is a $m$-term $N$ of the following form, then the reduction 
is local and consists of the rewriting $N\rightsquigarrow_{\beta}N'$ in $\mathcal{S}_x$ as follows: 
 \begin{center}
  \begin{tabular}{c} 
$\mathtt{postp}(\mathtt{connect\ to} (R))\rightsquigarrow_{\beta}\ []$.\\
$\mathtt{casel}~(N_0\wp N_1)\rightsquigarrow_{\beta}\ N_0;\qquad$%
$\mathtt{caser}~(N_0\wp N_1)\rightsquigarrow_{\beta}\ N_1$.\\
\end{tabular}
\end{center}
\noindent
If the principal operator of $\mathcal{R}ed$ is a binary $\mathtt{postpone}$ or $\mathtt{store}$, 
then the reduction is global and consists of the following rewriting. By the axioms in definiton 
\ref{acyclicity}, $\mathcal{R}ed$ occurs inside a computational context $\mathcal{S}_v$ in the rooted tree of nested $p$-terms of $\mathcal{S}_x$ and the rewriting takes place within $\mathcal{S}_v$. 

\medskip

\item If $\mathcal{R}ed$ has the form  
$\mathtt{postp}(\mathtt{z}\mapsto N, \mathtt{mkc}(M, \mathtt{y}))$,
then $\mathcal{S}_v$ is partitioned as 
$$
\mathcal{S}_v\ =\ \mathcal{R}ed \cup 
\mathcal{S}_{vyz}\bigl[y := \mathtt{y}(M),\ z:= \mathtt{z}(\mathtt{mkc}(M,\mathtt{y}))\bigr]
$$
\emph{(a simultaneous substitution of $\mathtt{y}(M)$ for $y$ and of 
$\mathtt{z}(\mathtt{mkc}(M,\mathtt{y}))$ for $z$ in $\mathcal{S}_{vyz}$).}
Then a {\em reduction} of $\mathcal{R}ed$ transforms the computational context as follows:
$$
\mathcal{S}_v\ =\  \mathcal{S}_{vyz}\bigl[y := N[z:= M],\ z:= M\bigr].
$$
\item If  $\mathcal{R}ed$ is a term with principal operator $\mathtt{store}$, then $\mathcal{S}_v$ 
is partitioned as $\mathcal{S}^-_{v} \cup \mathbf{S}$ where
\[
\mathbf{S}\ =\ \mathtt{store}(\overline{P}, N_1,\ldots, N_n, \mathtt{y}_1,\ldots, \mathtt{y}_n, \mathtt{z}, N) \]
\emph{Here $N$ is either $[M]$ or $\mathtt{connect\ to} (R)$ or $[[M_0][M_1]]$. } 

\medskip

\end{enumerate}

\noindent By the axioms \ref{acyclicity} $\overline{P}, N_1,\ldots, N_n$ form a computational context 
$\mathcal{S}_z$. 
Moreover $\mathtt{y}_1(\mathtt{z}([M]))$, $\ldots$, $\mathtt{y}_n(\mathtt{z}([M]))$ may occur in
 $\mathcal{S}^-_v$ so we write $\mathcal{S}^-_v$ as 
$\mathcal{S}^-_{y_1\ldots y_n v}\bigl[y_1 := \mathtt{y}_1(\mathtt{z}(M)),\ldots, 
y_n := \mathtt{y}_n(\mathtt{z}(M))\bigr]$.
\begin{itemize}
\item  If $N = [M]$, then 
\begin{displaymath}
  \mathcal{S}_v\ \rightsquigarrow_{\beta} 
  \mathcal{S}^-_{y_1\ldots y_n v}\bigl[y_1 := N_1[z:= M],\ldots, y_n := N_n[z:= M]\bigr]  
  \cup \{\overline{P}[z:= M]
\end{displaymath}
\item   If $N = \mathtt{connect\ to}(R)$, where $R$ belongs to $\mathcal{S}^-_v$, then 
   \[
   \mathcal{S}_v\ \rightsquigarrow_{\beta} \ 
   \mathcal{S}^-_{y_1\ldots y_n v}\bigl[y_1 := \mathtt{connect\ to}(R),\ldots, y_n := 
   \mathtt{connect\ to}(R)\bigr]
   \]
\item    If $N = [M_0, M_1]$, then 
   \begin{displaymath}
   \begin{array}{rl}
    \mathcal{S}_v\ \rightsquigarrow_{\beta}\ &  \mathcal{S}^-_{y_1\ldots y_n v} 
      \Bigl[y_1:= \bigl[ [\mathtt{y}_1(\mathtt{z}(M_0))], [\mathtt{y}_1(\mathtt{z}(M_1)]\bigr],\ldots\\ 
&\qquad\qquad
\ldots, y_n:= \bigl[ [\mathtt{y}_n(\mathtt{z}(M_0))], [\mathtt{y}_n(\mathtt{z}(M_1))]\bigr]\Bigr] \\
&       \cup  \{\mathtt{store}(\overline{N}, \overline{\mathtt{y}}, \mathtt{z}, M_0), 
       \mathtt{store}(\overline{N}, \overline{\mathtt{y}}, \mathtt{z}, M_1) \}
   \end{array}
   \end{displaymath}
   \end{itemize}
\end{defi}

\begin{rems}Here are some informal explanations about our calculus and notations.

\medskip

\noindent
(i) In our ``distributed'' model of computation  a redex arises when an $m$-term $M$ becomes part of another term; the rewriting of the redex has global effect. On the contrary, a $p$-term can only 
be sub-term of a $p$-term of the form $\mathtt{store}$, but such nesting has only a structural significance; no redex is created in this way. A  $p$-term $P$ sits in the ``control area'', waiting to become active as a redex if a suitable $m$-term is substituted inside it. A term $\mathtt{y}(M)$ denotes a variable $y$ that has become bound because of an operation in which the term $M$ is active; in some sense $\mathtt{y}(M)$ is an input which is no longer accessible. 
Later in the computation such an input may become active again in a term $R$ and ready for 
a substitution by a $m$-term $N$ in a rewriting of the form 
$R[y:= \mathtt{y}(M)] \rightsquigarrow R[y:= N]$.

\medskip

\noindent
(ii) When a $p$-term 
$\mathtt{store}(\overline{P}, N_1,\ldots, N_n, \mathtt{y}_1,\ldots, \mathtt{y}_n, \mathtt{z}, N)$ is created, the $m$-terms  $N_1, \ldots, N_n$ are set aside, together with  the local $p$-terms $P_1, \ldots P_m$, within the new $\mathtt{store}$ terms sitting in the control area, but the ``guarding terms'' $\mathtt{y}_1, \ldots, \mathtt{y}_n$ associated with $N_1, \ldots, N_n$ remain active, since they are part of other terms
in the context. Also the free variable $z$ occurring in the terms $P_j$ and $N_i$ becomes inaccessible and is substituted with $\mathtt{z}(N)$. Only the term $N$ is active in the storage operation. 
If $N = [M]$ then the computation is reactivated in $m$-terms $N_i[z:= M]$ and the guarding terms 
$\mathtt{y}_i$ are destroyed.
If $N = [M_0, M_1]$ then both of the ``guarding terms'' and the $\mathtt{store}$ term are copied; 
if  $N = \mathtt{connect\ to} (R)$ then the stored terms and the guarding terms are destroyed and replaced by pointers to the term $R$.

\medskip

\noindent
(iii) A term ``{\em make-coroutine}'' $\mathtt{mkc}(M, \mathtt{y})$ jumps from the 
term $M$ to an input $y$ which becomes inaccessible and thus is substituted by a term 
$\mathtt{y}(M)$ throughout the computational context. On the other hand 
a ``{\em postpone}'' term $\mathtt{postp}(\mathtt{z}\mapsto N, M')$ stores some 
{\em threads} of the computation from $z$ to $N$ (possibly a list of terms). As a consequence 
the input $z$ becomes inaccessible and is substituted by a term $\mathtt{z}(M')$ 
throughout the computational context. If $M'$ is $\mathtt{mkc}(M, \mathtt{y})$, then we
can reactivate the stored threads $N$ and free the variables $z$ and $y$ in the computational
context. The variable $z$ is substituted by $M$ wherever it occurs, i.e., as ${\xi}[z:= M]$. 
Moreover the threads $N$ are connected to $M$ through the substitution $N[z:= M]$ and 
the variable $y$ is substituted by $N[z:= M]$. Here we see a calculus with binding and substitution implemented as ``global effects'' in a co-intutionistic calculus through terms originally conceived by 
Tristan Crolard \cite{Crol04} as extensions of the $\lambda\mu$ calculus.
\end{rems} 
The term assignment to {\bf co-ILL} in sequent-style Natural Deduction notation is given in tables
\ref{nat-assign-mult} and \ref{nat-assign-exp}. 
Sequents are of the form $$x:E\triangleright \overline{P}\ |\ \overline{M}: \Gamma$$
where   
\begin{itemize}
\item the area of the succedent to the left of ``$|$'' may be called {\em ``control area''};  
\item $\overline{P} = P_1, \ldots, P_m$ is a sequence of p-terms;  
\item $\overline{M}: \Gamma$ stands for $M_1: C_1, \ldots, M_n: C_n$, where $\Gamma$ = 
$C_1, \ldots, C_n$;
\item if $\overline{R} = R_1, \ldots, R_n$ then $\overline{R}[x := N]$ stands for $R_1[x := N], \ldots, R_n[x := N]$.  
\item We shall also use the abbreviation $\kappa: \Gamma$ for $\overline{P} \ |\ \overline{M}: \Gamma$. 
If also  $\zeta : \Delta$ stands for $\overline{Q} \ |\ \overline{N}: \Delta$, then 
$\kappa: \Gamma, \zeta: \Delta$ stands for 
$\overline{P}, \overline{Q}\ |\ \overline{M}: \Gamma, \overline{N}: \Delta$.
\end{itemize}
\begin{table}[ht]
 \begin{center}
  \begin{small}
   \begin{tabular}{|c|}
   \hline
\\
\AxiomC{\it axiom}
\noLine
\UnaryInfC{$x: A\triangleright\ |\ x: A$}
\DisplayProof
 \qquad
\AxiomC{\it cut}
\noLine
\UnaryInfC{$v: E\triangleright \overline{P}\ |\ \overline{M} : \Gamma, M: A\qquad 
x: A\triangleright \overline{Q}\ |\  \overline{N}: \Delta$}
\UnaryInfC{$x: E\triangleright\overline{P}, \overline{Q}'[x:= M]\ |\ \overline{M}:\Gamma, \overline{N}[x:= M] :\Delta$}
\DisplayProof\\
\\
\AxiomC{$\bot$-intro}  
\noLine
\UnaryInfC{$x: E \triangleright \overline{P}\ |\ \overline{M} : \Gamma\qquad
(R \in \overline{P}\cup \overline{M})$}
\UnaryInfC{$x: E \triangleright \overline{P}\ |\ \overline{M}: \Gamma, \mathtt{connect\ to}(R) : \bot$}
\DisplayProof
\qquad
\AxiomC{$\bot$-elim}
\noLine
\UnaryInfC{$x: \bot\triangleright \mathtt{postp}(x) \ |$}
\DisplayProof\\ 
\\
We write $\overline{Q}'$ for $\overline{Q}[x:= \mathtt{x}(M)]$ and 
$\overline{N}'$ for $\overline{N}[x:= \mathtt{x}(M)]$\\
 \\
\AxiomC{$\smallsetminus$-intro }
\noLine
\UnaryInfC{$v:E\triangleright \overline{P}\ |\ \overline{M}: \Gamma, M: C \qquad  
x: D\triangleright \overline{Q}\ |\ \overline{N}: \Delta$}
\UnaryInfC{$v:E\triangleright \overline{P},\ \overline{Q}'\ |\ \overline{M}: \Gamma, 
\overline{N}': \Delta, \mathtt{mkc}(M, \mathtt{x}): C\smallsetminus D$}
\DisplayProof\\
\\
\AxiomC{$\smallsetminus$-elim}
\noLine
\UnaryInfC{$z: E \triangleright \overline{P}\ |\ \overline{M}: \Gamma, M: C\smallsetminus D\qquad
 x: C \triangleright \overline{Q}\ |\ N : D, \overline{N}: \Delta$}
\UnaryInfC{$z: E  \triangleright \overline{P},\ \overline{Q}',   
\mathtt{postp}(x\mapsto M, z)\ |\ \overline{M}:\Gamma, \overline{N}': \Delta$}
\DisplayProof
\\
\\
\AxiomC{$\wp$-intro}
\noLine
\UnaryInfC{$x: E\triangleright\overline{P}\ |\  \overline{M}: \Gamma, M_0: C_0,\  M_1: C_1$}
\UnaryInfC{$x: E\triangleright \overline{P}\ |\ \overline{M}: \Gamma, M_0\wp M_1: C_0\wp C_1$}
\DisplayProof\\
\\
\AxiomC{$\wp$-elim}
\noLine
\UnaryInfC{$z: E \triangleright \overline{Q}\ |\ \overline{N} : \Delta, N: C_0\wp C_1\hskip2in$} 
\noLine
\UnaryInfC{$\qquad\vdots\qquad\qquad  x_0: C_0 \triangleright \overline{P}_0 \ |\ \overline{M}_0:  \Gamma_0\qquad x_1: C_1 \triangleright \overline{P}_1 \ |\ \overline{M}_1: \Gamma_1$}
\UnaryInfC{$z: E\triangleright \overline{Q},\ \overline{P}'_0,\ \overline{P}'_1\ |\ 
\overline{N}: \Delta,\ \overline{M}'_0: \Gamma_0,\ \overline{M}'_1: \Gamma_1$}
\noLine
\UnaryInfC{ where $\overline{P}'_0 = \overline{P}[x_0:= \mathtt{casel}(z)], 
\overline{M}'_0 = \overline{M}_0[x_0:= \mathtt{casel}(z)],  $}
\noLine
\UnaryInfC{$\qquad \overline{P}'_1 = \overline{P}_1[x_0:= \mathtt{caser}(z)], 
\overline{M}'_1 = \overline{M}_1[x_0:= \mathtt{caser}(z)]$}
\DisplayProof\\
\\
\hline
  \end{tabular} 
  \end{small}
  \caption{Decorated Natural Deduction for {\bf multiplicative co-ILL}}
  \label{nat-assign-mult}
 \end{center} 
\end{table}
\begin{table}[ht]
 \begin{center}
  \begin{small}
   \begin{tabular}{|c|}
   \hline
\\
\AxiomC{\em dereliction}
\noLine
\UnaryInfC{$x: E\triangleright \kappa: \Gamma, M: C$}
\UnaryInfC{$x: E\triangleright \kappa: \Gamma, [\, M\,]: ?C$}
\DisplayProof
\\
\\
\qquad
\AxiomC{\em weakening}
\noLine 
\UnaryInfC{$x: E\triangleright \kappa: \Gamma$}
\UnaryInfC{$x: E\triangleright \kappa: \Gamma, \mathtt{connect\ to }(R): ?C$}
\noLine
\UnaryInfC{where $R \in \kappa$.} 
\DisplayProof
\qquad
\AxiomC{\em contraction}
\noLine 
\UnaryInfC{$x: E\triangleright \kappa: \Gamma, M: ?C, N: ?C$}
\UnaryInfC{$x: E\triangleright \kappa: \Gamma, [\,M, N\,] : ?C$}
\DisplayProof\\
 \\
\quad
\AxiomC{\em storage}
\noLine
\UnaryInfC{$z: E\triangleright \overline{P}\ |\ \overline{M}:\Gamma, M: ? C\qquad
x: C\triangleright \overline{Q}\ |\ \overline{N}:\, ?\Delta$}
\UnaryInfC{$z: E \triangleright \overline{P}, 
\mathtt{store}(\overline{Q}, \overline{N}, \overline{\mathtt{y}}, \mathtt{x}, N) \ |\ 
\overline{M}: \Gamma,  \overline{[\mathtt{y}(\mathtt{x}(N))]}:\, ?\Delta$}
\noLine
\UnaryInfC{where $\overline{N} = N_1, \ldots, N_m$ and  
$\overline{\mathtt{y}} = \mathtt{y}_1, \ldots, \mathtt{y}_m$}
\noLine
\UnaryInfC{and $\overline{[\mathtt{y}(\mathtt{x}(N))]} = 
\mathtt{y}_1(\mathtt{x}(N)), \ldots, \mathtt{y}_m(\mathtt{x}(N))$}
\DisplayProof\\
\\
\hline
  \end{tabular} 
  \end{small}
  \caption{Decorated Natural Deduction for {\bf co-ILL exponential}}
  \label{nat-assign-exp}
 \end{center} 
\end{table}

\subsection{Examples of multiplicative contexts.}\hfill\medskip

\noindent 1. The following computational context $\mathcal{S}_x$
\[
\mathcal{S}_x = \mathtt{postp}(x)\ \|\ \mathtt{connect\ to}(\mathtt{postp}(x))
\]
is correct. It is typed as follows: 
\begin{center}
\AxiomC{$\bot$-elim}
\noLine
\UnaryInfC{$x: \bot \triangleright \mathtt{postp}(x)$}
\RightLabel{$\bot$ intro}
\UnaryInfC{$x: \bot \triangleright \mathtt{postp}(x)\ |\  \mathtt{connect\ to}(\mathtt{postp}(x)) : \bot$}
\DisplayProof
\end{center}
This derivation may be regarded as the $\eta$-expansion of the axiom 
\[
x : \bot \triangleright x: \bot .
\]

\medskip

\noindent
2. Given the computational contexts $\mathtt{S}_x =  x\ \|\ \mathtt{connect\ to}(x)$ and 
$\mathcal{S}_y$ = $\mathtt{postp}(y)$, we obtain a correct  computational context by substitution of 
$\mathtt{connect\ to}(x)$ for $y$ in $\mathcal{S}_y$: 
\[
\mathtt{S}'_x =  x\ \|\ \mathtt{postp}(\mathtt{connect\ to}(x))
\]
$\mathtt{S}'_x$  $\beta$-reduces to 
$\mathtt{S}''_x =  x\ \|\ []$.

\medskip

\noindent
3. The following context $\mathcal{S}'_z$ is not correct: it violates Axiom 2 in definition \ref{acyclicity}.
\[
\mathcal{S}'_z =  \mathtt{postp}(y\mapsto [\mathtt{x}(y),\mathtt{x}(z)], \mathtt{mkc}(z, \mathtt{x}))\ 
\| \ \mathtt{mkc}(\mathtt{y}(t), \mathtt{x})
\]
Here we write $t$ for $\mathtt{mkc}(z, \mathtt{x})$.  The following context $\mathcal{S}_z$ is correct 
\[
\mathcal{S}_z = \mathtt{postp}(y\mapsto \mathtt{x}(y), \mathtt{mkc}(z, \mathtt{x}))\ \| \ 
\mathtt{mkc}(\mathtt{y}(t), \mathtt{x})\ \| \ \mathtt{x}(z) 
\]
and is typed as follows: 
\begin{small}
\begin{center}
\AxiomC{$z: C \triangleright z: C$}
\AxiomC{$x: C \triangleright x: C$}
\RightLabel{$\smallsetminus$-I}
\BinaryInfC{$z : C \triangleright \mathtt{mkc}(z, \mathtt{x}): C\smallsetminus C, \mathtt{x}(z) : C$}
\AxiomC{$y: C \triangleright y: C$}
\AxiomC{$x: C \triangleright x: C$}
\RightLabel{$\smallsetminus$-I}
\BinaryInfC{$y: C \triangleright \mathtt{mkc}(y, \mathtt{x}) : C\smallsetminus C, \mathtt{x}(y) : C$}
\RightLabel{$\smallsetminus$-E}
\BinaryInfC{$z : C \triangleright  \mathtt{postp}(y\mapsto \mathtt{x}(y), \mathtt{mkc}(z, \mathtt{x}))\ |\ 
 \mathtt{mkc}(\mathtt{y}(t), \mathtt{x}): C\smallsetminus C,  \mathtt{x}(z) : C$}
\DisplayProof
\end{center}
\end{small}
One could check that the above derivation is dual to the derivation 
\[
f: A \rightarrow A, x: A\ \triangleright\ (\lambda x.fx) x : A
\]
in the simply typed $\lambda$-calculus. 
A more substantial example of computation in our typed dual linear calculus is given in Appendix, 
Section \ref{example}. 
\section{Motivations: a probabilistic interpretation.}\label{ProbabilisticInterpretation}
In our setting co-intuitionistic logic admits a simple \emph{probabilistic interpretation} which fits well 
in the view of co-intuitionism as a logic of hypotheses. Indeed if co-intuitionistic logic is about the justification properties of hypotheses, then the co-intuitionistic consequence relation must be about the 
\emph{preservation of probability assignments} from the premise to the conclusions; a term calculus for such a logic must allow us to compute probabilities and verify the preservation property. We sketch 
our result only for the \emph{multiplicative linear} fragment, i.e.,
for typing derivations in the linear system with \emph{subtraction} and \emph{par} only.

\medskip

\noindent
We find it easier to state our result for a \emph{decorated sequent calculus} for multiplicative 
co-intuitionistic linear logic. Such a calculus is equivalent to our system of decorated sequent-style natural deduction; in fact its \emph{right} rules coincide with the \emph{introduction rules} and using \emph{cut} the \emph{left} rules given below are shown to be equivalent to the \emph{elimination rules.} 

\begin{defi}\label{prob-ass-constr}
To the judgements of linear co-intuitionistic logic we assign \emph{events in a probabilistic setting}. We write $\overline{\mathbf{C}}$, $\mathbf{C\cap D}$ and $\mathbf{C\cup D}$ for complementation, intersection and union between events; there is an impossible event $\emptyset$ and a certain event 
$\overline{\emptyset}$. A \emph{probabilistic assignment} is a map 
$(\ )^P : \mathbf{judg} \rightarrow \mathbf{events}$ satisfying the following constraints: 
\begin{center}
\begin{tabular}{rl} 
if $(M : C)^P = \mathbf{C}$ and $(x: D)^P = \mathbf{D}$, then 
& $(\mathtt{mkc}(M, x) : C \smallsetminus D)^P = \mathbf{C}\cap \overline{{\bf D}}$;\\
if $(M_0: C_0)^P = \mathbf{C}_0$ and $(M_1: C_1)^P =  \mathbf{C}_1$, then 
& $(M_0 \wp M_1: C_0 \wp C_1)^P = \mathbf{C}_0\cup \mathbf{C}_1$;\\
if $(M : C_0 \wp C_1)^P =  \mathbf{C}$ then & $(\mathtt{casel}(M): C_0)^P \subseteq \mathbf{C}$\\
and & $(\mathtt{caser}(M) : C_1)^P  \subseteq \mathbf{C}$;\\
$(\mathtt{postp}(M))^P = \emptyset $ and & $(\mathtt{postp}(x\mapsto M, N))^P = \emptyset$.
\end{tabular}
\end{center}
\end{defi} 

\begin{prop} \emph{ ({\bf Decomposition property}) }\label{Decomposition}
Let $d$ be a sequent calculus derivation of $x: H\triangleright t_1:
C_1, \ldots, t_n: C_n$ and let $(\ )^P : \mathbf{judg} \rightarrow
\mathbf{events}$ be an assignment to the judgements in $d$ satisfying the constraints of
Definition \ref{prob-ass-constr}, and suppose
$\mathbf{H}, \mathbf{C}_1, \ldots, \mathbf{C}_n$ are assigned to $x: H \triangleright t_1: C_1,\dots,t_n :
C_n$.  There are pairwise disjoint events $\mathbf{C}'_1\subseteq\mathbf{C}_1,\dots,
\mathbf{C}'_n\subseteq\mathbf{C}_n$ such that 
\[\left(\mathbf{C}'_1\cup\cdots\cup\mathbf{C}'_n\right)\cap\mathbf{H}=\mathbf{H}.
\]
The events $\mathbf{C}'_1,\dots,\mathbf{C}'_n$ can be constructed
from the dependencies of the terms $t_1,\dots,t_n$.
\end{prop}

\begin{exa}\label{E:new}
  Consider the following very simple example:
\begin{center}
\AxiomC{$x : C\triangleright x : C$}
\AxiomC{$y : D\triangleright y : D$}
\RightLabel{}
\BinaryInfC{$x:C\triangleright\mathtt{mkc}(x,y):C\smallsetminus
  D,\mathtt y(x):D$}
\DisplayProof\\
\end{center}
  If the event $\mathbf C$ is assigned to $x:C$, and $\mathbf D$ is
  assigned to
  $y:D$, but also to $\mathtt y(x):D$, then we have the following inclusions
\begin{center}
\AxiomC{$\mathbf C\subseteq\mathbf C$}
\AxiomC{$\mathbf D\subseteq\mathbf D$}
\RightLabel{}
\BinaryInfC{$\mathbf C\subseteq(\mathbf C\cap\overline{\mathbf D})\cup
  \mathbf{D}$}
\DisplayProof\\
\end{center}
  We have equality only by assigning $\mathbf C\cap\mathbf D$ to
  $\mathtt y(x):D$, as suggested by the dependency of $\mathtt y(x):D$
  on $x:C$.
 \begin{center}
\AxiomC{$\mathbf C = \mathbf C$}
\AxiomC{$\mathbf D = \mathbf D$}
\RightLabel{}
\BinaryInfC{$\mathbf C = (\mathbf C\cap\overline{\mathbf D})\cup
 (\mathbf{C} \cap \mathbf{D})$}
\DisplayProof 
\end{center} 
\end{exa}

\proof By induction on $d$. The case of assumptions $x: H \triangleright x: H$ is obvious and that of
cut is immediate from the inductive hypothesis.

\medskip

\noindent
\emph{Subtraction right:} by inductive hypothesis we may assume that the assignments to the premises
$v:E\triangleright \kappa: \Gamma, M: C$ and $x: D\triangleright \zeta: \Delta$ satisfy the conditions of
the lemma, i.e., that  $\mathbf{((\bigcup\Gamma)\cup C)\cap E = E}$ and 
$\mathbf{(\bigcup \Delta)\cap D = D}$, where the events in 
$\mathbf{\Gamma}$ are pairwise disjoint and so are those in $\mathbf{\Delta}$. 
In the term assignment to the conclusion
\[
v:E\triangleright \kappa: \Gamma, \mathtt{mkc}(M, \mathtt{x}): C\smallsetminus D, 
\zeta[x:= \mathtt{x}(M)]: \Delta
\]
in all terms $\zeta: \Delta$ the variable $x : D$ has been replaced by $\mathtt{x}(M)$, where $M : C$. 
We interpret this fact as \emph{the instruction that in the context of the conclusion the disjoint events
$\mathbf{D'}_j \subseteq \mathbf{D}_j$ must be $\mathbf{D}_j \cap (\mathbf{C\cap D})$} 
 for each $D_j \in \Delta$. 
Then 
\[
\mathbf{C = (C\cap\overline{\mathbf{D}}) \cup  (C\cap D)} =  
(\mathbf{C}\cap \overline{\mathbf{D}}) \cup \bigl(\mathbf{C \cap D} \cap \mathbf{\bigcup \Delta}\bigr) = 
(\mathbf{C}\cap \overline{\mathbf{D}}) \cup (\cup_j \mathbf{D'}_j)
\]
hence $\mathbf{C\cap E = [(C\cap \overline{\mathbf{D}})\cap E] \cup [(\cup_j \mathbf{D'}_j)\cap E]}$. 
Thus 
\[
\mathbf{((\cup_i C_i) \cup (C\cap\overline{D}) \cup (\cup_j D'_j)) \cap E = E}.
\]
\begin{center}
\begin{tabular}{|c|}
\hline
\\
\AxiomC{\emph{subtraction}-L}
\noLine
\UnaryInfC{$x: C \triangleright \overline{P}\ |\ M : D, \overline{M}: \Delta$}
\UnaryInfC{$z: C\smallsetminus D  \triangleright\ \overline{P}[x:= \mathtt{x}(z)],   
\mathtt{postp}(x\mapsto M, z)\ |\ \overline{M}[x:= \mathtt{x}(z)]: \Delta$}
\DisplayProof \\
\\
\hline
\end{tabular}
\end{center}
\noindent
\emph{Subtraction left:}  suppose that by inductive hypothesis we have an assignment to the premise
$x: C \triangleright \overline{P}\ |\ M: D, \overline{M}: \Delta$, thus 
$\mathbf{D}$ and $\mathbf{D_i \in \Delta}$ are pairwise disjoint and 
$\mathbf{(D \cup (\bigcup \Delta)) \cap C = C}$, then obviously 
$\mathbf{(\bigcup \Delta) \cap C\cap \overline{\mathbf{D}} = C\cap \overline{\mathbf{D}}}$.
Clearly p-terms have empty assignment.
\begin{center}
\begin{tabular}{|c|}
\hline
\\
\AxiomC{\emph{par}-L}
\noLine
\UnaryInfC{$x_0: C_0 \triangleright \overline{P}_0 \ |\ \overline{M}_0:  \Gamma_0\qquad 
x_1: C_1 \triangleright \overline{P}_1 \ |\ \overline{M}_1: \Gamma_1$}
\UnaryInfC{$z: C_0 \wp C_1 \triangleright \overline{P}'_0,\ \overline{P}'_1\ |\ 
\overline{M}'_0: \Gamma_0,\ \overline{M}'_1: \Gamma_1$}
\noLine
\UnaryInfC{ where $\overline{P}'_0 = \overline{P}[x_0:= \mathtt{casel}(z)], 
\overline{M}'_0 = \overline{M}_0[x_0:= \mathtt{casel}(z)]$}
\noLine
\UnaryInfC{$\qquad \overline{P}'_1 = \overline{P}_1[x_0:= \mathtt{caser}(z)], 
\overline{M}'_1 = \overline{M}_1[x_0:= \mathtt{caser}(z)]$}
\DisplayProof\\
\\
\hline
\end{tabular}
\end{center}
\noindent
In the case of \emph{par right} there is nothing to prove; in the case of \emph{par left} we only need to 
make sure that the events $\mathbf{C}_0$ and $\mathbf{C}_1$ assigned to $C_0$ and $C_1$ are disjoint. If not, assign $\mathbf{C}_0$ to $C_0$ and $\mathbf{C}_1 \cap \overline{\mathbf{C_0}}$
to $C_1$, or alternatively $\mathbf{C}_0 \cap \overline{\mathbf{C}}_1$ to $C_0$ and $\mathbf{C}_1$
to $C_1$. \qed
\begin{rems}
(i) Since events are assigned to expressions $t : X$ rather than to formulas $X$,  
if $t : X$ and $u : X$ occur in the same context then $(t: X)^P$ and $(u: X)^P$ are events 
that may or may not be disjoint of each other. 

\medskip

\noindent
(ii) The \emph{common sense} reading of the co-intuitionistic consequence relation 
$H \vdash C_1, \ldots, C_n$ is as follows.
\begin{quotation}
\noindent
\emph{If it is justified to make the hypothesis $H$, then it is justified to make the hypotheses 
$C_1, \ldots, C_n$.}
\end{quotation}
The probabilistic interpretation gives a mathematical counterpart of this reading. 
\begin{quotation}
\noindent
\emph{If the probability of the event $\mathbf{H}$ assigned to $x: H$ is greater than zero, then the conditional probability of the union of the events $\mathbf{C}_1, \ldots, \mathbf{C}_n$ assigned to 
$t_1: C_1, \ldots, t_n: C_n$, given $\mathbf{H}$ is equal to one.}
\end{quotation}
The indexing of the terms $t_1, \ldots, t_n$ can be regarded as computational devices for  
verifying such an interpretation in the sense of the Decomposition Property.
\end{rems}  
\section{Categorical Semantics}
We recall the definition of a symmetric monoidal category.

\begin{defi}
A {\em symmetric monoidal category (SMC)} 
$(\mathbb{C}, \bullet, 1, \alpha, \lambda, \rho, \gamma)$, 
 is a category $\mathbb{C}$ equipped with a bifunctor 
 $\bullet : \mathbb{C}\times \mathbb{C} \rightarrow \mathbb{C}$ with a neutral element $1$ and natural isomorphisms $\alpha, \lambda, \rho$ and $\gamma$: 
\begin{enumerate}
\item $\alpha_{A,B,C,}: A\bullet (B\bullet C) \stackrel{\sim}{\longrightarrow} (A\bullet B)\bullet C$;
\item $\lambda_A: 1 \bullet A \stackrel{\sim}{\longrightarrow } A$
\item $\rho_A: A \bullet 1  \stackrel{\sim}{\longrightarrow } A$
\item $\gamma_{A,B}: A \bullet B \stackrel{\sim}{\longrightarrow } B \bullet A$.
\end{enumerate}
which satisfy the following coherence diagrams.
\begin{center}
$\xymatrix@R=.4in@C+20 pt{A \bullet (B \bullet (C \bullet D)) \ar[d]_{id_A\bullet\alpha_{B,C,D}}
\ar[r]^{\alpha_{A,B,C\bullet D}} &
(A\bullet B)\bullet(C \bullet D)\ar[r]^{\alpha_{A\bullet B, C, D}}&(((A\bullet B)\bullet C)\bullet D\\
A\bullet ((B\bullet C)\bullet D)  \ar[rr]_{\alpha_{A, B\bullet C,D}}&   & 
(A\bullet (B\bullet C)) \bullet D \ar[u]_{\alpha_{A,B,C}\bullet id_D}\\}$ \\
$\xymatrix@R=.4in@C+10 pt{(A \bullet B) \bullet C \ar[d]_{\gamma_{A,B}\bullet id_C} \ar[r]^{\alpha^{-1}_{A,B,C}} &
A\bullet (B\bullet C)\ar[r]^{\gamma_{A, B\bullet C}}&(B\bullet C)\bullet A \ar[d]^{\alpha^{-1}_{B,C,A}}\\
(B\bullet A)\bullet C \ar[r]_{\alpha^{-1}_{B,A,C}}&  B\bullet (A\bullet C) \ar[r]_{id_B\bullet\gamma_{A,C}} & 
B\bullet (C\bullet A)\\}$ \\
$\xymatrix@R=.4in{A \bullet (1 \bullet B) \ar[dr]_{id_A\bullet \lambda_B} \ar[rr]^{\alpha_{A,1,B}} &&
(A\bullet 1)\bullet B \ar[dl]_{\rho_A\bullet id_B}\\
&A\bullet B&\\} 
\qquad 
$\xymatrix@R=.4in{A \bullet B\ar[dr]_{id_{A\bullet B}} \ar[r]^{\gamma_{A,B}} & 
B\bullet A \ar[d]^{\gamma_{B,A}}\\
& A\bullet B\\}
\\
$\xymatrix@R=.4in{A \bullet 1 \ar[dr]_{\rho_a} \ar[rr]^{\gamma_A, 1} &
& 1\bullet A \ar[dl]^{\lambda_A}\\
&A&\\}$
\end{center}
The following equality is also required to hold: 
$\lambda_{1} = \rho_{1} : 1 \bullet 1 \rightarrow 1$.
\end{defi}

\medskip

\noindent
Given a {\em signature} $Sg$, consisting of a collection of types $\sigma_i$ and a collection of 
{\em sorted function symbols} $f_j : \sigma_1, \ldots, \sigma_n \rightarrow \tau$ and given 
a SMC $(\mathbb{C}, \bullet, 1, \alpha, \lambda, \rho, \gamma)$,  a {\em structure} $\mathcal{M}$
for $Sg$ is an assignment of an object $[\![\sigma]\!]$ of $\mathbb{C}$ for each type $\sigma$ and of 
a morphism $[\![f]\!] : [\![\sigma_1]\!]\bullet\ldots\bullet[\![\sigma_n]\!]\rightarrow [\![\tau]\!]$ for each function $f : \sigma_1, \ldots, \sigma_n \rightarrow \tau$ of $Sg$.

\medskip

\noindent
The types of terms in context $\Delta = [M_1: \sigma_1, \ldots, M_n: \sigma_n]$ are 
interpreted as $[\![\sigma_1, \sigma_2, \ldots, \sigma_n]\!]$ =   
$(\ldots ([\![\sigma_1]\!]\bullet[\![\sigma_2]\!])\ldots ) \bullet [\![\sigma_n]\!]$; left associativity is also 
intended for concatenations of type sequences $\Gamma, \Delta$. Thus we need the following functions 
$\mathtt{Split}(\Gamma, \Delta): [\![\Gamma, \Delta]\!] \rightarrow [\![\Gamma]\!]\bullet [\![\Delta]\!]$ 
\begin{displaymath}
\mathtt{Split}(\Gamma, \Delta) \left\{ \begin{array}{ll}
\lambda^{-1}_{\Delta} & \hbox{if}\ \Gamma = \emptyset\\
\rho^{-1}_{\Gamma}   & \hbox{if}\ \Delta = \emptyset\\
id_{\Gamma\bullet A} &  \hbox{if}\ \Delta = A\\
\mathtt{Split}(\Gamma, \Delta')\bullet id_A ; \alpha^{-1}_{\Gamma,\Delta',A} &\hbox{if}\
\Delta =\Delta', A.
\end{array}\right.  
\end{displaymath}
and 
$\mathtt{Join}(\Gamma, \Delta):[\![\Gamma]\!]\bullet [\![\Delta]\!] \rightarrow [\![\Gamma, \Delta]\!]$
\begin{displaymath}
\mathtt{Join}(\Gamma, \Delta) \left\{ \begin{array}{ll}
\lambda_{\Delta} & \hbox{if}\ \Gamma = \emptyset\\
\rho_{\Gamma}   & \hbox{if}\ \Delta = \emptyset\\
id_{\Gamma\bullet A} &  \hbox{if}\ \Delta = A\\
\alpha_{\Gamma,\Delta',A} ; \mathtt{Join}(\Gamma, \Delta')\bullet id_A ; &\hbox{if}\ 
\Delta =\Delta', A.
\end{array}\right.  
\end{displaymath} 
Similarly we have $\mathtt{Split}_n(\Gamma_1,\ldots,\Gamma_n): 
[\![\Gamma_1,\ldots,\Gamma_n]\!]\rightarrow[\![\Gamma_1]\!]\bullet\ldots\bullet[\![\Gamma_n]\!]$ 
and $\mathtt{Join}_n$.

\medskip

\noindent
The semantics of terms in context is then specified by induction on terms: 
\begin{center}
\begin{tabular}{c}
$[\![x:\sigma\triangleright x:\sigma]\!] =_{df} id_{[\![\sigma]\!]}$\\
\\
$[\![x:\sigma\triangleright f(M_1,\ldots, M_n): \tau]\!] =_{df} [\![x:\sigma\triangleright M_1:\sigma_1]\!]\bullet
\ldots\bullet [\![x:\sigma\triangleright M_n:\sigma_n]\!]; [\![ f ]\!]$
\end{tabular}
\end{center}   
The {\em Exchange right} rule is handled implicitly by symmetry in the model 
(see \cite{Bierman94}, Lemma 13): 
$$
[\![x:\sigma \triangleright \overline{M}:\Gamma, N: \tau, M:\sigma]\!] = 
[\![x:\sigma \triangleright \overline{M}:\Gamma, M:\sigma, N: \tau]\!]; 
\alpha^{-1}_{\Gamma,\sigma, \tau}; id_{\Gamma}\bullet \gamma_{\sigma, \tau}; 
\alpha_{\Gamma,\tau, \sigma}
$$
(Notice that, as a notational convenience, we are sometimes reusing the names of types to denote their interpretation as objects). 
One then proves by induction on the type derivation that substitution in the term calculus 
corresponds to composition in the category (\cite{Bierman94}, Lemma 13): 
\begin{lem} Let $x:\sigma\triangleright \overline{M}: \Gamma, M: \tau$ and 
$y:\tau\triangleright \overline{N}: \Delta$ be derivable terms in context, then 
\begin{center}
\begin{tabular}{rl}
$[\![x:\sigma \triangleright \overline{M}:\Gamma, \overline{N}[y:= M]\, \Delta ]\!]$ =&\\ 
$[\![x:\sigma \triangleright \overline{M}:\Gamma, M: \tau]\!];$& 
$id_{\Gamma}\bullet [\![y:\tau \triangleright \overline{N}: \Delta]\!]; \mathtt{Join}(\Gamma,\Delta)$ 
\end{tabular}
\end{center}
\end{lem}\medskip

\noindent Let $M$ be a structure for a signature $Sg$ in a SMC $\mathbb{C}$. Given an equation in
context for $Sg$ 
$$x:\sigma\triangleright \overline{M}:\Gamma, M = N : \tau$$
we say that the structure {\em satisfies} the equation if the morphisms assigned to 
$x:\sigma\triangleright \overline{M}:\Gamma, M : \tau$ and to 
$x:\sigma\triangleright \overline{M}:\Gamma, N : \tau$ are equal. Then given an algebraic theory 
$Th = (Sg, Ax)$, a structure $\mathcal{M}$ for $Sg$ is a {\em model} for $Th$ if it satisfies all the 
axioms in $Ax$. 

\medskip

\begin{lem} Let $\mathbb{C}$ be a SMC, $Th$ an algebraic theory and $\mathcal{M}$ a model of
$Th$ in $\mathbb{C}$. Then $\mathcal{M}$ satisfies the equations in context in Table \ref{basiceqn}.
\end{lem}

\begin{table}[htpb]
 \begin{center}
  {\footnotesize
   \begin{tabular}{|c|}
\hline
\\
\AxiomC{$z:D\triangleright \overline{M}:\Gamma, M:\sigma$}
\RightLabel{\em Refl}
\UnaryInfC{$z:D\triangleright \overline{M}:\Gamma, M = M:\sigma$}
\DisplayProof
\quad
\AxiomC{$z:D\triangleright \overline{M}:\Gamma, M = N:\sigma$}
\RightLabel{\em Symm}
\UnaryInfC{$z:D\triangleright \overline{M}:\Gamma, N = M:\sigma$}
\DisplayProof\\
\\
\AxiomC{$z:D\triangleright \overline{M}:\Gamma, M_0=M_1:\sigma$}
\AxiomC{$z:D\triangleright \overline{M}:\Gamma, M_1=M_2:\sigma$}
\RightLabel{\em Trans}
\BinaryInfC{$z:D\triangleright \overline{M}:\Gamma, M_0 = M_2:\sigma$}
\DisplayProof\\
\\
\AxiomC{$z:D\triangleright \overline{M}:\Gamma, M_0=M_1:\sigma$}
\AxiomC{$x:\sigma\triangleright \overline{N}:\Delta,  N_0=N_1:\tau$}
\RightLabel{\em Subst}
\BinaryInfC{$z:D\triangleright \overline{M}:\Gamma, \overline{N}[x:= M_0] = \overline{N}[x:= M_1] : \Delta, N_0[x:= M_0] = N_1[x:= M_1]:\tau$}
\DisplayProof\\
\\
\hline
  \end{tabular}}
    \caption{}
     \label{basiceqn}
    \end{center} 
\end{table}
\subsection{Analysis of the rules of co-intuitionistic linear logic}
We work with symmetric monoidal categories satisfying {\em the dual condition to closure}, 
namely, with monoidal categories of the form $(\mathbb{C}, \bullet, \smallsetminus, 1, \alpha, \lambda, \rho, \gamma)$ such that for all objects $A$ in $\mathbb{C}$, the functor $A \bullet - $ has a 
{\em left} adjoint $- \smallsetminus A$. We call such monoidal categories {\em left closed}. 

\medskip

\noindent
Given a symmetric monoidal category $\mathbb{C}$, its opposite is also symmetric monoidal. 
If $\mathbb{C}$ is closed, i.e., $A \bullet -$ has a right adjoint, then certainly $\mathbb{C}^{op}$
has a left adjoint. It is well-known that in a symmetric monoidal closed category $\mathbb{C}$
we can construct a model of {\em multiplicative intuitionistic linear logic}, hence it is certainly not
surprising that a model of multiplicative co-intuitionistic linear logic may be constructed in 
$\mathbb{C}^{op}$. The point of the exercise that follows, however, is to check that the dual 
linear calculus given above in Section \ref{ConjecturalTermAssign} is indeed suitable for the 
construction of such an interpretation. We consider the rules for each connective in turn.

\subsection{Linear disjunction Par}\label{lindisj} 
\ref{lindisj}.1. {\em Par introduction.} 
The introduction rule for Par is of the form 
\begin{center} 
\AxiomC{$x:D\triangleright \kappa:\Theta, M_0:A, M_1:B$}
\RightLabel{$\wp$ I}
\UnaryInfC{$x:D\triangleright \kappa:\Theta, M_0\,\wp\, M_1: A\wp B$}
\DisplayProof
\end{center}
This suggests an operation on Hom-sets of the form 
$$
\Phi_{D, \Theta}: \mathbb{C}(D, \Theta\bullet A\bullet B) \rightarrow 
\mathbb{C}(D, \Theta\bullet A\wp B)
$$
{\em natural in} $\Theta$ and $D$. Given $e: D\rightarrow \Theta\bullet A\bullet B$, $d: D'\rightarrow D$
and $h: \Theta\rightarrow\Theta'$, naturality yields 
$$
\Phi_{D', \Theta'}(d; e; h\bullet id_A\bullet id_B) = d; \Phi_{D, \Theta}(e); h\bullet id_{A\wp B}
$$
In particular, letting $e = id_{\Theta} \bullet id_A \bullet id_B$, $d: D \rightarrow \Theta\bullet A\bullet B$ and $h = id_{\Theta}$ we have 
$$
\Phi_{D, \Theta}(d) = d; \Phi_{\Theta} (id_{\Theta}\bullet id_A \bullet id_B)
$$
By functoriality of $\bullet$ we have $id_{A}\bullet id_{B} = id_{A\bullet B}$. Hence, writing  
{\bf PAR} for $\Phi_{\Theta} (id_{\Theta}\bullet id_{A \bullet B})$ we have 
$\Phi_{D, \Theta}(d) = d; \mathbf{PAR}$. We define
$$
[\![x:D \triangleright \kappa :\Theta, M\wp N: A\wp B]\!] =_{df} 
     [\![x:D \triangleright \kappa: \Theta, M:A, N:B]\!]; \mathbf{PAR}.
$$

\vspace{3ex}

\noindent
\ref{lindisj}.2. {\em Par elimination}. The Par elimination rule has the form 
\begin{center} 
\AxiomC{$z:D\triangleright \kappa:\, \Upsilon, N:A\,\wp\, B$}
\AxiomC{$x:A\triangleright \zeta:\, \Gamma$}
\AxiomC{$y:B\triangleright \xi:\, \Delta$}
\RightLabel{$\wp$ E}
\TrinaryInfC{$z:D\triangleright \kappa:\, \Upsilon, \zeta[x:= \mathtt{casel}\,N]:\Gamma, 
\xi[y:=\mathtt{caser}\,N]:\Delta$}
\DisplayProof
\end{center}
This suggests an operation on Hom-sets of the form 
$$
\Psi_{D,\Upsilon,\Gamma,\Delta}: \mathbb{C}(D, \Upsilon\bullet A\wp B)\times
\mathbb{C}(A, \Gamma)\times\mathbb{C}(B, \Delta)\rightarrow 
\mathbb{C}(D, \Upsilon\,\bullet\, \Gamma\bullet \Delta)
$$
{\em natural in} $D,\Upsilon,\Gamma,\Delta$. Given morphisms $g:D\rightarrow \Upsilon\bullet A\wp B$, 
$e: A\rightarrow\Gamma$ and $f: B\rightarrow \Delta$ and also $a: D'\rightarrow D$, 
$p:\Upsilon\rightarrow \Upsilon'$, $c: \Gamma\rightarrow \Gamma'$ and $d: \Delta \rightarrow \Delta'$ naturality yields
\begin{center}
\begin{tabular}{rl}
$\Psi_{D',\Upsilon',\Gamma',\Delta'}\bigl((a;g;p\bullet id_{A\wp B}), (e;c),$ & $(f;d)\bigr)$ =\\ 
$a; \Psi_{D,\Upsilon,\Gamma,\Delta}(g,e,f);$ & 
$p\bullet c\bullet d; \mathtt{Join}(\Upsilon',\Gamma',\Delta').$
\end{tabular}
\end{center}
In particular, setting $e = id_A$, $f = id_B$ and also $a = id_D$, $p = id_{\Upsilon}$, we get 
$$
\Psi_{D,\Upsilon,\Gamma,\Delta}(g, c, d) = \Psi_{D,\Upsilon,\Gamma,\Delta}(g, id_A, id_B); 
id_{\Upsilon}\bullet c\bullet d; \mathtt{Join}(\Upsilon,\Gamma, \Delta) 
$$
Writing $(g)^{\ast}$ for $\Psi_{D,\Upsilon}(g, id_A, id_B)$ we define
\begin{center}
\begin{tabular}{l}
$[\![z:D\triangleright\kappa:\, \Upsilon, \zeta[x:= \mathtt{casel}\ N]:\, \Gamma, \xi[y:= \mathtt{caser}\,N]:\,  \Delta]\!]=_{df}$\\
$[\![z:D\triangleright \kappa:\, \Upsilon, N:\, A\,\wp\, B]\!]^*; 
id_{\Upsilon}\bullet [\![x:A\triangleright \zeta:\, \Gamma]\!]\bullet[\![y:B\triangleright\xi:\, \Delta]\!];
\mathtt{Join}(\Upsilon, \Gamma, \Delta)$.\\
\end{tabular}
\end{center}

\vspace{3ex}

\noindent
\ref{lindisj}.3. {\em Equations in context.} We have equations in context of the form
\begin{equation}\label{context:par}
\framebox{\quad\AxiomC{\bf $\wp - \beta$ rules:}
\noLine
\UnaryInfC{$z:D \triangleright \kappa:\,\Theta, M_0:\, A, M_1:\, B\qquad x:A \triangleright \zeta:\,\Gamma\quad y:B \triangleright \xi:\,\Delta$}
\UnaryInfC{$z:D \triangleright \kappa:\, \Theta, 
\zeta[x:= \mathtt{casel}\, (M_0\,\wp\,M_1)] = \zeta[x:= M_0] : \Gamma$}
\noLine
\UnaryInfC{\strut}
\noLine
\UnaryInfC{$z:D \triangleright \kappa:\,\Theta, M_0:\, A, M_1:\, B\qquad x:A \triangleright \zeta:\,\Gamma\quad y:B \triangleright \xi:\,\Delta$}
\UnaryInfC{$z:D \triangleright \kappa:\,\Theta,  
\xi[y:= \mathtt{caser}\, (M_0\wp M_1)] = \xi[y:= M_1] :\Delta$}
\DisplayProof}
\end{equation}
Let $q: D\rightarrow \Theta\bullet A\bullet B$, $m:A\rightarrow\Gamma$ and $n: B\rightarrow\Delta$.
Then to satisfy the above equations in context we need that the following diagram commutes:
\begin{center}
$\xymatrix@R=.4in{D\ar[r]^q & \Theta\bullet A\bullet B \ar[d]_{\wp} \ar[r]^{id_{\Theta}\bullet m\bullet n} &
\Theta\bullet\Gamma\bullet\Delta\\
&\Theta\bullet A\wp B \ar[r]_{\ast} &\Theta\bullet A\bullet B \ar[u]_{id_{\Theta}\bullet m\bullet n}\\}$
\end{center}   
We make the assumption that the above decomposition is unique. 
Moreover, supposing $\Theta$ empty and $m = id_A$, $n = id_B$, $q = id_A\bullet id_B = id_{A\bullet B}$ 
we obtain $(id_A\bullet id_B; \mathbf{PAR})^{\ast} = id_A\bullet id_B$ and similarly 
$(id_{A\,\wp\, B})*; \mathbf{PAR} = id_{A\,\wp\, B}$; hence we may conclude that there is a natural isomorphism 
\begin{center}
\AxiomC{$D \rightarrow \Gamma\bullet A\bullet B$}
\doubleLine
\UnaryInfC{$D \rightarrow \Gamma\bullet A\wp B$}
\DisplayProof
\end{center}
so we can identify $\bullet$ and $\wp$. Finally we see that the following $\eta$ equation in context 
are also satisfied:  
\begin{equation}\label{eta:par}
\quad\framebox{
\AxiomC{$z:D \triangleright \kappa: \Upsilon, M:A\,\wp\, B$}
\AxiomC{\bf $\wp - \eta$ rule:}
\noLine
\UnaryInfC{$x:A\triangleright x:A$}
\AxiomC{$y:B\triangleright y:B$}
\TrinaryInfC{$z:D\triangleright \kappa:\Upsilon, \mathtt{casel}(M)\ \wp\ \mathtt{caser}(M) = M:\, A\,\wp\, B $}
\noLine
\UnaryInfC{\strut}
\DisplayProof\quad}
\end{equation}
\subsection{Linear subtraction}\label{linsubtr}
\ref{linsubtr}.1. {\em Subtraction introduction.} The introduction rule for subtraction has the form 
\begin{center} 
\AxiomC{$x:D\triangleright \kappa:\,\Gamma, M:\,A$}
\AxiomC{$y:B\triangleright \zeta:\,\Delta$}
\RightLabel{$\smallsetminus$ I}
\BinaryInfC{$x:D\triangleright \kappa:\,\Gamma, 
\zeta[y := \mathtt{y}(M)]: \Delta, \mathtt{mkc}(M, \mathtt{y}): A\smallsetminus B$}
\DisplayProof
\end{center}
This suggests a natural transformation with components
$$
\Phi_{D, \Gamma,\Delta}: \mathbb{C}(D, \Gamma\bullet A)\times\mathbb{C}(B, \Delta) \rightarrow 
\mathbb{C}(D, \Gamma\bullet \Delta\bullet A\smallsetminus B)
$$
natural in $D, \Gamma, \Delta$. Taking morphisms $e: D\rightarrow \Gamma\bullet A$, 
$f:B\rightarrow \Delta$ and $a: D'\rightarrow D$, $c:\Gamma\rightarrow \Gamma'$, 
$d: \Delta\rightarrow \Delta'$, by naturality we have
$$
\Phi_{D', \Gamma',\Delta'}\left((a; e; c\bullet id_A), (f;d)\right) = 
a; \Phi_{D, \Gamma,\Delta}(e, f); c\bullet d\bullet id_{A\smallsetminus B}; 
\mathtt{Join}(\Gamma',\Delta', A\smallsetminus B)
$$
In particular, taking $a=id_D$, $c=id_{\Gamma}$, $d: B\rightarrow\Delta$ and $f= id_B$ we have:
$$
\Phi_{D, \Gamma,\Delta}(e, d) = \Phi_{D,\Gamma} (e, id_B); 
id_{\Gamma}\bullet d \bullet id_{A\smallsetminus B}; \mathtt{Join}(\Gamma, \Delta, A\smallsetminus B)
$$
Writing $\mathbf{MKC}^B_{D,\Gamma}(e)$ for $ \Phi_{D,\Gamma} (e, id_B)$, 
$\Phi_{D, \Gamma,\Delta}(e, d)$ can be expressed as the composition 
$$
\mathbf{MKC}^B_{D,\Gamma}(e); id_{\Gamma}\bullet d\bullet id_{A\smallsetminus B}
$$ 
where $\mathbf{MKC}^B_{D,\Gamma}$ is a natural transformation with components 
$$
\mathbf{MKC}^B_{D, \Gamma}: \mathbb{C}(D, \Gamma\bullet A)\times\mathbb{C}(B, B) \rightarrow 
\mathbb{C}(D, \Gamma\bullet B \bullet A\smallsetminus B)
$$
so we make the definition 
\begin{center}
\begin{tabular}{l}
$[\![x:D \triangleright \kappa:\,\Gamma, \zeta[y:= \mathtt{y}(M)], 
\mathtt{mkc}(M,\mathtt{y}):\, A\smallsetminus B]\!] =_{df}$\\
\quad $\mathbf{MKC}^B_{D,\Gamma}[\![x:D\triangleright\kappa:\Gamma, M:A]\!];
id_{\Gamma}\bullet [\![y:B\triangleright \zeta:\Delta]\!]\bullet id_{A\smallsetminus B}; 
\mathtt{Join}(\Gamma,\Delta, A\smallsetminus B)$\\
\end{tabular}
\end{center}

\vspace{3ex}

\noindent
\ref{linsubtr}.2. {\em Subtraction elimination.} The subtraction elimination rule has the form 
\begin{center} 
\AxiomC{$x:D\triangleright \kappa:\Gamma, M:A\smallsetminus B$}
\AxiomC{$y:A\triangleright \xi:\Delta, N:B$}
\RightLabel{$\smallsetminus$ E}
\BinaryInfC{$x:D\triangleright  \mathtt{postp}(y\mapsto N, M), \kappa:\Gamma, 
\xi[y:= \mathtt{y}(M)]:\Delta$} 
\DisplayProof
\end{center}
This suggests a natural transformation with components
$$
\Psi_{D, \Gamma,\Delta}: \mathbb{C}(D, \Gamma\bullet (A\smallsetminus B))\times
\mathbb{C}(A, \Delta\bullet B) \rightarrow 
\mathbb{C}(D, 1\bullet \Gamma\bullet \Delta)
$$
natural in $D, \Gamma, \Delta$. Given $e: D\rightarrow \Gamma\bullet (A\smallsetminus B)$, 
$f: A\rightarrow \Delta\bullet B$ and also $a: D'\rightarrow D$, $c: \Gamma\rightarrow\Gamma'$, 
$d:\Delta\rightarrow\Delta'$, naturality yields 
$$
\Psi_{D', \Gamma',\Delta'}\left((a; e; c\bullet id_{A\smallsetminus B}), (f;d\bullet id_B)\right) = 
a; \Psi_{D, \Gamma,\Delta}(e, f); c\bullet d; \mathtt{Join}(\Gamma',\Delta')
$$
In particular, taking $a: D\rightarrow \Gamma\bullet(A\smallsetminus B)$, 
$e = id_{\Gamma\bullet(A\smallsetminus B)}$, $c= id_{\Gamma}$, $d: id_{\Delta}$, we obtain
$$
\Psi_{D, \Gamma,\Delta}(a, f) = a; \Psi_{D, \Gamma,\Delta}(id_{\Gamma\bullet(A\smallsetminus B)}, f); \mathtt{Join}(\Gamma,\Delta)
$$
Writing $\mathbf{POSTP}(f)$ for $\Phi_{D, \Gamma,\Delta}(id_{\Gamma\bullet(A\smallsetminus B)}, f)$
we define 
\begin{center}
\begin{tabular}{l}
$[\![x:D \triangleright \mathtt{postp}(y\mapsto N, M), \kappa :\Gamma, \xi[y:= \mathtt{Y}(M)]\!] =_{df}$\\
\quad $[\![x:D\triangleright \kappa:\Gamma, M: A\smallsetminus B]\!]; id_{\Gamma}\bullet
\mathbf{POSTP}[\![y:A \triangleright \xi: \Delta, N: B]\!]; \mathtt{Join}(\Gamma,\Delta)$ 
\end{tabular}
\end{center}

\medskip

\noindent
\ref{linsubtr}.3. {\em Equations in context.}  We have equations in context of the form 
\begin{equation}\label{context:subtr}
\framebox{\small
\AxiomC{$x: D\triangleright \overline{M}:\Gamma, M:A$} 
\AxiomC{\bf $\smallsetminus - \beta$ rule:}
\noLine
\UnaryInfC{$y: B\triangleright \overline{N}:\Delta$} 
\AxiomC{$z: A\triangleright \overline{L}:\Lambda, L:B$} 
\TrinaryInfC{$x:D\triangleright \overline{M}:\Gamma, \left[\overline{N}':\Delta, \mathbf{red}
\overline{L}': \Lambda\right] = \left[\overline{N}[y:= L[z::= M]\,], \overline{L}[z:= M]\right]$}
\noLine
\UnaryInfC{\strut}
\DisplayProof}
\end{equation}
where $\overline{N}' = \overline{N}[y:= Y(M)]$, 
$\mathbf{red} = \mathtt{postp}\left(z\mapsto L, \mathtt{mkc}(M,Y)\right)$ and 
$ \overline{L}' = \overline{L}[z:= \mathtt{mkc}(M,Y)]$.

\medskip

\noindent
Given morphisms $n: D\rightarrow \Gamma\bullet A$ and $m: A\rightarrow \Delta\bullet B$, for these equations to be satisfied we need the following diagram to commute:
\begin{center}
$\xymatrix@R=.4in{D\ar[d]_{\mathbf{MKC}^B(n)} \ar[rr]^n &\qquad & 
\Gamma\bullet A\ar[d]^{id_{\Gamma}\bullet m}\\
\Gamma\bullet(A\smallsetminus B)\bullet B\ar[rr]_{\mathbf{POSTP}(m)\bullet id_B} & &\Gamma\bullet\Delta\bullet B\\
}$
\end{center}
in particular, taking $n = id_A$ we have 
\begin{center}
$\xymatrix@R=.4in{A \ar[d]_{\mathbf{MKC}^B(id_A)}\ar[r]^m&\Delta\bullet B\\
(A\smallsetminus B)\bullet B \ar[ur]|{\mathbf{POSTP}(m)\bullet id_B} &\\
}$
\end{center}
Assuming the above decomposition to be unique, we can show that the $\eta$ equation in context
is also satisfied:     
\begin{equation}\label{eta:subtr}
\framebox{\small 
\AxiomC{$z:D \triangleright \overline{N}: \Delta, M: A\smallsetminus B$}
\AxiomC{\bf $\smallsetminus - \eta$ rule}
\noLine
\UnaryInfC{$x:A \triangleright x:A$}
\AxiomC{$y:B \triangleright y:B$}
\TrinaryInfC{$z:D \triangleright \overline{N}:\Delta, \bigl[\mathtt{mkc}(X(M), Y): A\smallsetminus B,
\mathtt{postp}(x\mapsto Y(x), M) 
\bigr] = M: A\smallsetminus B$}
\noLine
\UnaryInfC{\strut}
\DisplayProof}
\end{equation}
and conclude that there is a natural isomorphism between the maps 
\begin{center}
\AxiomC{$A \rightarrow \Delta\bullet B$}
\doubleLine
\UnaryInfC{$A\smallsetminus B\rightarrow \Delta$}
\DisplayProof
\end{center}
i.e., that $\smallsetminus$ is the left adjoint to the bifunctor $\bullet$.  
\subsection{Unit}\label{units}
\ref{units}.1 {\em Unit rules.}  The introduction and elimination rules for the unit $\bot$ are
\begin{center}
\begin{tabular}{cc} 
\AxiomC{$\bot$ {\em introduction}}
\noLine
\UnaryInfC{$x:D\triangleright \kappa:\,\Gamma$}
\UnaryInfC{$x:D\triangleright \kappa:\,\Gamma, \mathtt{connect\ to}(R):\,\bot$}
\noLine
\UnaryInfC{where $R \in \kappa$.}
\DisplayProof \quad
&\quad 
\AxiomC{$\bot$ {\em elimination}}
\noLine
\UnaryInfC{$x:\bot \triangleright \mathtt{postp}(x)$}
\noLine
\UnaryInfC{\strut}
\DisplayProof
\end{tabular} 
\end{center}
The elimination rule is interpreted by a unique map $\langle\rangle: \bot \rightarrow 1$.

\medskip

\noindent 
The introduction rule requires a natural transformation with components
\begin{center}
$\Phi_{D,\Gamma}: \mathbb{C}(D, \Gamma)\rightarrow \mathbb{C}(D, \Gamma\bullet \bot)$
\end{center}
natural in $D$ and $\Gamma$. Given morphisms $e:D\rightarrow \Gamma$, $d:D'\rightarrow D$ 
and $c:\Gamma\rightarrow\Gamma'$, naturality yields
\begin{center}
$\Phi_{D', \Gamma'}(d;e;c) = d;\Phi_{D, \Gamma}(e);c$.
\end{center} 
Letting $d:D\rightarrow\Gamma$ and $e = id_{\Gamma}$, $c = id_{\Gamma\bullet\bot}$ we have 
$$\Phi_{D, \Gamma}(d) = d;\mathbf{Bot}_{\Gamma}$$
where we write $\mathbf{Bot}_{\Gamma}$ for $\Phi_{\Gamma}(id_{\Gamma})$.
We define 
$$
[\![x:D\triangleright \kappa:\,\Gamma, \mathtt{connect\ to}(x):\bot]\!] =_{df} 
[\![x:D \triangleright \kappa:\,\Gamma]\!];\mathbf{Bot}_{\Gamma}. 
$$
\ref{units}.2. {\em Equations in context.}
We may assume the operation $\mathbf{Bot}_{\Gamma}$ to be compatible with the generalized associativity and commutativity properties of $\bullet$, so that for $\Gamma = C_1, \ldots, C_n$ we have  
$$
\Phi_{C_1,\ldots C_i, \bot, \ldots, C_n}(id_{\Gamma}) : 
C_1\bullet \cdots\bullet C_i\bullet \bot,\bullet\cdots C_n\quad = \quad
\Phi_{\Gamma}(id_{\Gamma}) : C_1\bullet\cdots\bullet C_n\bullet \bot
$$
for all $i \leq n$. Together with naturality of $\mathbf{Bot}_{\Gamma}$ these yield the equations
in context
\begin{equation}
\framebox{$
\begin{tabular}{c}\label{rewiring}
\AxiomC{$x:D\triangleright \kappa:\,\Gamma$}
\UnaryInfC{$x:D\triangleright \kappa:\,\Gamma, \bigl[\mathtt{connect\ to}(R_i) = \mathtt{connect\ to}(R_j)\bigr]$}
\noLine
\UnaryInfC{$R_i, R_j \in \kappa$}
\DisplayProof\\
\\
\AxiomC{$x:D\triangleright \kappa:\,\Gamma\quad(R_i\in \kappa)$}
\UnaryInfC{$y:E\triangleright \zeta:\,\Gamma'\quad(R_j \in \zeta)$}
\UnaryInfC{$y:E\triangleright \zeta, \bigl[\mathtt{connect\ to}(R_i) = \mathtt{connect\ to}(R_j)\bigr]$}
\DisplayProof
\end{tabular}$} 
\end{equation}
that correspond to the {\em rewiring properties} of $\bot$-links in the proof-net representation by 
\cite{BCS,BCST}. Moreover the equation in context        
\begin{equation}\label{context:bot}
\framebox{\quad
\AxiomC{\bf $\bot - \beta$ rule}
\noLine
\UnaryInfC{$x:D\triangleright \kappa:\,\Gamma\quad 
y:\bot\triangleright\mathtt{postp}(y)\qquad (R \in \kappa)$}
\UnaryInfC{$x:D\triangleright [\kappa:\, \Gamma, 
\mathtt{postp} (\mathtt{connect\ to}(R)\, ) 
= \kappa:\,\Gamma]$}
\DisplayProof
\quad}
\end{equation}
requires that for any $m: D\rightarrow \Gamma$ the following diagram commutes: 
\begin{center}
$\xymatrix@R=.4in{D \ar[r]^{m;\mathbf{Bot}} \ar[dr]_m& 
\Gamma\bullet\bot\ar[d]^{id_{\Gamma}\bullet \langle\rangle;\lambda_{A}}\\
&\Gamma.\\}$
\end{center}
Assuming that this decomposition is unique and taking $m = id_A$ we have that 
$\mathbf{Bot}_A; id_A\bullet\langle\rangle;\lambda_A = id_A$. Arguing as before, 
we see that there is a natural isomorphism 
\begin{center}
\AxiomC{$D\rightarrow\Gamma\bullet 1$}
\doubleLine
\UnaryInfC{$D\rightarrow\Gamma\bullet \bot$}
\DisplayProof
\end{center}
(so we identify $\bot$ and 1) and that the following equation in context is satisfied:
\begin{equation}\label{eta:bot}
\quad\framebox{
\AxiomC{\bf $\bot - \eta$ rule:}
\noLine
\UnaryInfC{$z:D \triangleright \kappa:\, \Gamma, M:\bot\hskip1in x:\bot\triangleright \mathtt{postp}(x)$}
\UnaryInfC{$z:D\triangleright \kappa:\, \Gamma, \bigl[\mathtt{connect\ to}(\mathtt{postp}(M)\, ):\bot 
= M : \bot\bigr]$}
\noLine
\UnaryInfC{\strut}
\DisplayProof
\quad}
\end{equation} 
Let $\mathcal{L}$ be the signature having 
\begin{itemize}
\item the types given by the following grammar on a collection of ground types $\gamma$: 
$$
A\ :=\ \gamma\ |\ \bot\ |\ A\wp A\ |\ A\smallsetminus A
$$  
\item a collection of sorted function symbols including $\mathtt{connect\ to}(-)$, 
$\mathtt{postp}(-)$,  
$\wp(-, -)$,  
$\mathtt{casel}(-)$,   
$\mathtt{caser}(-)$,  
$\mathtt{mkc}(-,-)$,  
$\mathtt{postp}(-,-)$.
\end{itemize}
We have proved the following
\begin{thm}
Let $\mathcal{T} = (\mathcal{L}, \mathcal{A})$ be a theory with signature $\mathcal{L}$ having as axioms the equations in context in Table \ref{basiceqn} and in (\ref{context:par}) - 
(\ref{eta:bot}). Let $(\mathbb{C}, \bullet, 1, \smallsetminus, \alpha, \lambda, \rho,\gamma)$ be a symmetric monoidal left-closed category and $\mathcal{M}$ a structure 
for $\mathcal{L}$ in $\mathbb{C}$. Then $\mathcal{M}$ satisfies the equations in $\mathcal{A}$.
\end{thm}  
Moreover, define the {\em syntactic category} as the category $\mathcal{C}$ which has the formulas of multiplicative co-intuitionistic linear logic as objects and typed terms of the form 
$x:E \triangleleft \kappa:\Gamma$ ({\em modulo} renaming of the variable $x$) as morphisms. 
Set $x:E\triangleright \kappa: \Gamma\ =\ y:E\triangleright \zeta: \Gamma$
iff $\kappa = \zeta[y:= x]$ is derivable from equations in context  in Table \ref{basiceqn} and in (\ref{context:par}) - (\ref{eta:bot}). Then we have 
\begin{thm} The syntactic category is a symmetric monoidal left-closed category. 
\end{thm}  
From this fact the {\em categorical completeness} theorem follows. 

\section{Extension to co-intuitionistic linear logic with coproducts and exponential}
Let $\mathcal{L}^{\oplus}$ be $\mathcal{L}$ extended with additive disjunction $\oplus$ and 
the familiar functions $\mathtt{inl}: A \rightarrow A\oplus B$, $\mathtt{inr}: B\rightarrow A\oplus B$ and 
$\mathtt{case}: A\oplus B \times (A\rightarrow C) \times (B\rightarrow C)\rightarrow C$. 
Then it is easy to extend the above result to show that if $\mathbb{C}$ has also the structure of 
coproducts, then a structure for $\mathcal{L}^{\oplus}$ in $\mathbb{C}$ satisfies also the theory 
$\mathcal{T}^{\oplus}$ where $\mathcal{A}$ is extended with familiar equations in context for 
$\mathtt{inl}$, $\mathtt{inr}$ and $\mathtt{case}$. We shall not pursue this extension here. 

\vspace{3ex}

\noindent
The extension of $\mathcal{T}$ to a theory with the exponential {?} ({\em why not?}) is less simple.
On one hand, one can dualize Benton, Bierman, De Paiva, Hyland's definition of a linear category 
\cite{BBHdP92,Bierman94} and obtain in this way a sound and complete categorical semantics for 
co-intuitionistic linear logic. The construction of weakly distributive categories with storage operators 
based on proof-nets by Blute, Cockett and Seely \cite{BCS} provides a categorical model for both 
exponentials {\bf !} and {\bf ?}. On the other hand, the semantics for the exponential {\bf !} can recovered in the context of Nick Benton's treatment of {\em Linear Non Linear} logic \cite{Benton95}.
After dualizing the linear part of {\bf LNL} one should be able to recover the semantics for {\bf ?}
and at the same time obtain a framework where the duality of intuitionistic and co-intuitionistic
logic can be studied. We leave the development of this approach to future work and focus on the categorical semantics of the multiplicative and exponential {\bf ?} fragment of co-intuitionstic linear logic.   

\subsection{Co-intuitionistic linear categories} 
We begin by dualizing the definition of a linear category \cite{BBHdP92,Bierman94}.
\begin{defi} 
A {\em dual linear category} $\mathbb{C}$ consists of
\begin{enumerate}
\item A symmetric monoidal left-closed category together with
\item a symmetric co-monoidal monad $(?, \eta, \mu, \mathtt{n}_{-,-}, \mathtt{n}_{\bot})$ 
{\em (namely, the functor {\bf ?} is co-monoidal with respect to $\wp$ and the linear transformation 
$\eta, \mu$ are co-monoidal)}  such that 
\item[](i) - each free ?-algebra $(?A, \mu_A)$ carries naturally the structure of a commutative 
$\wp$-monoid {\em (i.e., for each $(?A, \mu_A)$ there are distinguished monoidal natural transformations $i_A: \bot\rightarrow ?A$ and $c_A: ?A\wp?A\rightarrow ?A$ which form 
a commutative monoid and are algebra morphisms)}; 
\item[](ii) - whenever $f: (?A, \mu_A) \rightarrow (?B, \mu_B)$ is a morphism of free algebras, then it is also a monoid morphism.
\end{enumerate}
\end{defi} 

\begin{rems}
By Maietti, Maneggia de Paiva and Ritter (see \cite{MaiettiManeggiaPaivaRitter}, Prop. 25), 
condition 2(ii) is equivalent to the requirement that $\mu$ is a monoidal morphism.\medskip

\noindent
(i) To say that the functor {\bf ?} is symmetric co-monoidal means that it comes equipped with a {\em comparison} natural transformation $\mathtt{n}_{A,B}: ?(A\wp B)\rightarrow ?A\wp ?B$ and a morphism
$\mathtt{n}_{\bot}: ?\bot \rightarrow \bot$, satisfying 
\begin{small}
\begin{center}
$\xymatrix@R=.4in{?(\bot\,\wp\,A)\ar[d]^{?\lambda_A}\ar[r]^{\mathtt{n}_{\bot,A}} &?\bot\,\wp\,?A
\ar[d]^{\mathtt{n}_{\bot}\,\wp\, id_{?A}}\\
?A &\ar[l]_{\lambda_{?A}}\ \bot\, \wp\, ?A \\}$
and 
$\xymatrix@R=.4in{?(A\,\wp\, \bot) \ar[d]^{?\rho_A} \ar[r]^{\mathtt{n}_{A, \bot}} &?A\,\wp\,?\bot
\ar[d]^{id_{?A}\ \wp\ \mathtt{n}_{\bot}}\\
?A &\ar[l]_{\rho_{?A}}\ ?A\, \wp\, \bot \\}$
\\
$\xymatrix@R=.4in{?((A\,\wp\,B)\wp\,C)\ar[r]^{\mathtt{n}_{A\,\wp,B,C}} \ar[d]_{?\alpha^{-1}_{A,B,C}}\ &\ 
           ?(A\,\wp\,B)\wp\,?C\ar[r]^{\mathtt{n}_{A,B}\,\wp\,id_C}\ &\ 
                      (?A\,\wp\,?B)\wp\,?C \ar[d]^{\alpha^{-1}_{?A,?B,?C}} \\
?(A\,\wp\,(B\wp\,C)) \ar[r]_{\mathtt{n}_{A,B\,\wp\,C}}\ &\     
           ?A\,\wp\,?(B\wp\,C) \ar[r]_{id_{?A}\,\wp\,\mathtt{n}_{B,C}}\ &\ 
                    ?A\,\wp\,(?B\wp\,?C) \\}$
\\
$\xymatrix@R=.4in{?(A\,\wp\,B)\ar[d]^{?\gamma_{A,B}}\ar[r]^{\mathtt{n}_{A,B}} 
            &?A\,\wp\,?B \ar[d]^{\gamma_{?A,?B}} \\
?(B\,\wp\,A) \ar[r]^{\mathtt{n}_{B,A}} & ?B\,\wp\,?A \\}$ and naturality: 
$\xymatrix@R=.4in{?(A\wp B) \ar[d]^{?(f\wp g)} \ar[r]^{\mathtt{n}_{A,B}} & ?A \wp ?B \ar[d]^{?f\wp ?g}\\
?(A'\wp B') \ar[r]^{\mathtt{n}_{A',B'}} & ?A' \wp ?B'\\}$ 
\end{center}
\end{small}

\noindent(ii) To say that $\eta$ and $\mu$ are co-monoidal is to say that the following diagrams commute:
\begin{small}
\begin{center}
$\xymatrix@R=.4in{A \wp B \ar[r]^{\eta_A\wp\eta_A} \ar[d]_{\eta_{A\wp B}} & ?A \wp ?B\\
?(A \wp B) \ar[ur]_{\mathtt{n}_{A,B}}\\}$ and 
$\xymatrix@R=.4in{\bot \ar[r]^{\eta_{\bot}} \ar[dr]_{id}&?\bot\ar[d]^{\mathtt{n}}\\
&\bot\\}$
\\
$\xymatrix@R=.4in{
??(A \wp B)\ar[r]^{\mu_{A\wp B}} \ar[d]_{?\mathtt{n}_{A,B}}&?(A\wp B)\ar[dd]^{\mathtt{n}_{A,B}}\\
?(?A \wp ?B) \ar[d]_{\mathtt{n}_{?A,?B}} & \\
??A\, \wp\, ??B \ar[r]_{\mu_A\wp \mu_B} & ?A\, \wp\, ?B \\}$ 
and
$\xymatrix@R=.4in{??\bot \ar[d]_{?\mathtt{n}_{\bot}} \ar[r]^{\mu_{\bot}}& ?\bot \ar[d]^{\mathtt{n}_{\bot}}\\
?\bot \ar[r]_{\mathtt{n}_{\bot}} & \bot\\}$
\end{center}
\end{small}

\noindent(iii) To say that the natural transformations $i_A: \bot\rightarrow ?A$ and 
$c_A: ?A\wp?A\rightarrow ?A$ are monoidal means that they are compatible with the 
comparison maps, i.e., that  the following diagrams commute:   
\begin{small}
\begin{center}
\begin{tabular}{c}
$\xymatrix@R=.4in{\bot \ar[r]^{id} \ar[d]_{i_{\bot}} &\bot\\
?\bot \ar[ur]_{\mathtt{n}_{\bot}}&\\}$
\quad
$\xymatrix@R=.4in{\bot\,\wp\,\bot \ar[r]^{\lambda=\rho}& \bot\\
?\bot\,\wp\,?\bot \ar[r]_{c_{\bot}}\ar[u]^{\mathtt{n}_{\bot}\,\wp\,\mathtt{n}_{\bot}}&
 ?\bot \ar[u]_{\mathtt{n}_{\bot}}\\}$
\quad
$\xymatrix@R=.4in{\bot\,\wp\,\bot \ar[d]_{\lambda = \rho}\ar[r]^{i_A\,\wp\,i_B} & ?A \wp ?B\\
\bot \ar[r]_{i_{A\wp B}} & ?(A\,\wp\,B) \ar[u]_{\mathtt{n}_{A,B}}\\}$\\
\end{tabular}
\end{center}
\end{small}
\begin{small}
\begin{center}
\begin{tabular}{c}
$\xymatrix@R=.4in{(?A\,\wp\,?A)\,\wp\,(?B\,\wp\,?B)\ar[rr]^{c_A\,\wp\,c_B} && ?A\,\wp\,?B\\
(?A\,\wp\,?B)\,\wp(?A\,\wp\,?B) \ar[u]^{iso} &\\
?(A\,\wp\,B)\,\wp\,?(A\,\wp\,B)\ar[u]^{\mathtt{n}_{A,B}\wp \mathtt{n}_{A,B}}\ar[rr]_{c_{A\wp B}} &&
	?(A\,\wp\,B)\ar[uu]_{\mathtt{n}_{A,B}}\\}$\\
\end{tabular}
\end{center}
\end{small}
where {\em iso} is the canonical isomorphism derived from symmetry and associativity;
\begin{small}
\begin{center}
\begin{tabular}{c}
$\xymatrix@R=.4in{\bot \ar[r]^{i_{A}} & ?A\\
?\bot \ar[u]^{\mathtt{n}_{\bot}} \ar[r]_{?i_A} &??A \ar[u]_{\mu_A}\\}$
\qquad and \qquad
$\xymatrix@R=.4in{?A\,\wp\,?A \ar[r]^{c_A} &?A\\
??A\,\wp\,??A \ar[u]^{\mu_A\,\wp\,\mu_A}&\\
?(?A\,\wp\,?A) \ar[u]^{\mathtt{n}_{?A,?A}} \ar[r]_{?c_A} & ??A \ar[uu]_{\mu_A}\\}$\\
\end{tabular}
\end{center}
\end{small}

\noindent(iv) Finally for the free algebra morphisms to be monoid morphism we require that the following
diagrams commute: 
\begin{small}
\begin{center}
\begin{tabular}{c}
$\xymatrix@R=.4in{
?A\,\wp\,?A \ar[r]^{c_A} & ?A\\
??A\,\wp\,??A \ar[u]^{\mu_A\,\wp\,\mu_A} \ar[r]_{c_{?A}} &??A\ar[u]_{\mu_A}\\}$
\qquad
$\xymatrix@R=.4in{\bot \ar[dr]_{i_{?A}} \ar[r]^{i_A} &?A\\
&??A\ar[u]_{\mu_A}\\
}$
\end{tabular}
\end{center}
\end{small}
\end{rems}

\subsection{Term and equations in context}
To sketch a proof that a dual linear category is a model of co-intuitionistic linear logic with storage
operator {\bf ?} we give the term in context and the equation in context relevant to the dereliction, 
weakening, contraction and storage rules. These conditions are dual to those in Figures 4.1-4.5 in 
G.~M.~Bierman's thesis \cite{Bierman94}, pp.~112-142. Since in our context the exponential rules
for dereliction, weakening and contraction do not involve {\tt let} constructions, some of these 
conditions result immediately from properties of substitution. 
 \begin{table}[ht]
 \begin{center}
  \begin{small}
   \begin{tabular}{|c|}
   \hline
\\
\AxiomC{$v:E\triangleright \kappa: \Gamma, M:\, ?C$}
\AxiomC{$x:C\triangleright \overline{Q}\ |\ \overline{N}:\, ?\Delta$}
\BinaryInfC{$v:E\triangleright \kappa:\Gamma, 
\mathtt{store}(\overline{Q}, \overline{N}, \overline{\mathtt{y}}, \mathtt{x}, M)\ |\ \overline{\mathtt{y}}(\mathtt{x}(M)):\,?\Delta$}
\DisplayProof\\
\\
\AxiomC{\em dereliction}
\noLine
\UnaryInfC{$x: E\triangleright \kappa: \Gamma, M: C$}
\UnaryInfC{$x: E\triangleright \kappa: \Gamma, [\, M\,]: ?C$}
\DisplayProof\\
\\
\AxiomC{\em weakening}
\noLine 
\UnaryInfC{$x: E\triangleright \kappa: \Gamma$}
\UnaryInfC{$x: E\triangleright \kappa: \Gamma, \mathtt{connect\ to }(R): ?C$}
\noLine
\UnaryInfC{where $R \in \kappa$.} 
\DisplayProof
\quad
\AxiomC{\em contraction}
\noLine 
\UnaryInfC{$x: E\vdash \kappa: \Gamma, M: ?C, N: ?C$}
\UnaryInfC{$x: E\vdash \kappa: \Gamma, [\,M, N\,] : ?C$}
\DisplayProof\\
\\
\hline
  \end{tabular} 
  \end{small}
  \caption{Term in context judgements for the {\bf ?} storage operator}
  \label{storage-termassign}
 \end{center} 
\end{table}
There are three Equations in Context expressing ``$\beta$ reductions'' for the storage operator
in Table \ref{storage-reductions}. 
\begin{table}[!ht]
 \begin{center}
  \begin{small}
   \begin{tabular}{|c|}
   \hline
\\
{\bf Dereliction - Storage:}
\\
\AxiomC{$v:E\triangleright \kappa: \Gamma, M: C$}
\AxiomC{$x:C\triangleright \overline{Q}\ |\ \overline{N}:\, ?\Delta$}
\BinaryInfC{$v:E\triangleright \kappa:\Gamma, 
\bigl[ \mathtt{store}(\overline{Q}, \overline{N}, \overline{\mathtt{y}}, \mathtt{x}, [M])\ |\ \overline{\mathtt{y}}(\mathtt{x}(\,[M]\,)):\,?\Delta = \bigr.$} 
\noLine
\UnaryInfC{$\hskip.6in \bigl. = \overline{Q}[x:= M]\ |\ \overline{N}[x:= M]:\, ?\Delta \bigr]$}
\DisplayProof\\
\\
{\bf Contraction - Storage:}
\\
\AxiomC{$v:E\triangleright \kappa: \Gamma, M_0:\, ?C, M_1:\, ?C$}
\AxiomC{$x:C\triangleright \overline{Q}\ |\ \overline{N}:\, ?\Delta$}
\BinaryInfC{$v:E\triangleright \kappa:\Gamma, 
\bigl[\mathtt{store}(\overline{Q}, \overline{N}, \overline{\mathtt{y}}, \mathtt{x}, [M_0, M_1])\ |\ 
\overline{\mathtt{y}}(\mathtt{x}(\,[M_0,M_1]\,)):\,?\Delta = \bigr.$} 
\noLine
\UnaryInfC{$\qquad\qquad =  
\mathtt{store}(\overline{Q}, \overline{N}, \overline{\mathtt{y}}, \mathtt{x}, M_0), 
\mathtt{store}(\overline{Q}, \overline{N}, \overline{\mathtt{y}}, \mathtt{x}, M_1])\ |$}
\noLine
\UnaryInfC{$\hskip1in\bigl. \langle[\overline{\mathtt{y}}(\mathtt{x}(M_0)), \overline{\mathtt{y}}(\mathtt{x}(M_1))]\rangle:\, ?\Delta \bigr]$}
\noLine
\UnaryInfC{where $?\Delta = ?D_1, \ldots ?D_m$ and  
$\langle[\overline{\mathtt{y}}(\mathtt{x}(M_0)),\overline{\mathtt{y}}(\mathtt{x}(M_1))]\rangle: ?\Delta$
stands for} 
\noLine
\UnaryInfC{$[\mathtt{y}_1(\mathtt{x}(M_0)), \mathtt{y}_1(\mathtt{x}(M_1))]:\ ?D_1, \ldots, 
[\mathtt{y}_m(\mathtt{x}(M_0)), \mathtt{y}_m(\mathtt{x}(M_1))]:\ ?D_m $}
\DisplayProof\\
\\
{\bf Weakening - Storage:}
\\
\AxiomC{$v:E\triangleright \kappa: \Gamma$}
\AxiomC{$x:C\triangleright \overline{Q}\ |\ \overline{N}:\, ?\Delta$}
\AxiomC{$R \in \kappa$}
\TrinaryInfC{$v:E\triangleright \kappa:\Gamma, \bigl[
\mathtt{store}( \overline{Q}, \overline{N}, \overline{\mathtt{y}}, \mathtt{x}, \mathtt{connect\ to}(R))\ |\ 
\bigr.$} 
\noLine
\UnaryInfC{$\hskip1in \overline{\mathtt{y}}(\mathtt{x}(\mathtt{connect\ to}(R)):\,?\Delta 
= \bigl. |\ \langle[\overline{\mathtt{connect\ to}}(R)]\rangle:\, ?\Delta \bigr]$}
\noLine
\UnaryInfC{where $?\Delta = ?D_1, \ldots ?D_m$ and   
$|\ \langle[\overline{\mathtt{connect\ to}}(R)]\rangle:\, ?\Delta$ stands for }
\noLine
\UnaryInfC{\hskip1in $|\ \mathtt{connect\ to}(R):\ ?D_1, \ldots, \mathtt{connect\ to}(R):\ ?D_m$}
\DisplayProof\\
\\
\hline
  \end{tabular} 
  \end{small}
  \caption{Equations in context for the {\bf ?} storage operator}
  \label{storage-reductions}
 \end{center} 
\end{table}
Finally there are {\em Categorical Equations in Context} in Table \ref{categorical-equincontext}.
\begin{table}[!h]
 \begin{center}
  \begin{small}
   \begin{tabular}{|c|}
   \hline
\\
{\bf Monad:} \\
\AxiomC{$z:?A\triangleright \bigl[ \mathtt{store}([\,[x]\,], \mathtt{y}, \mathtt{x}, z)\ |\ \mathtt{y}'(\mathtt{x}'(z)):\, ?A\ = x: ?A \bigr]$}
\DisplayProof\\
\\
{\bf Algebra 1}\\
\AxiomC{$v:E\triangleright \kappa: \Gamma, M: ?C$}
\AxiomC{$x:C\triangleright \overline{P}\ |\ \overline{N}: ?\Delta$}
\BinaryInfC{$v:E\triangleright \kappa: \Gamma, \overline{P}[x:= \mathtt{x}(M)],\hskip1in$} 
\noLine
\UnaryInfC{$ \bigl[\mathtt{store}(\langle\overline{N}, \mathtt{connect\ to}(R)\rangle, 
\langle\overline{\mathtt{y}}, \mathtt{y}\rangle, \mathtt{x}, M)\ |\ 
\overline{\mathtt{y}}(\mathtt{x}(M)):\, ?\Delta, \mathtt{y}(\mathtt{x}(M)):\, ?A = \bigr.$} 
\noLine
\UnaryInfC{$\bigl. = \mathtt{store}(\overline{N}, \overline{\mathtt{y}}, \mathtt{x}, M)\ |\ 
\overline{\mathtt{y}}(\mathtt{x}(M)):\, ?\Delta, \mathtt{connect\ to}(R'):\, ?A\bigr]$}
\noLine 
\UnaryInfC{where $R \in \overline{P}\cup \overline{N}$ and $R' \in \overline{P}[x:= \mathtt{x}(M)]\cup 
\overline{\mathtt{y}}(\mathtt{x}(M))$}
\DisplayProof\\
\\
{\bf Algebra 2}\\
\AxiomC{$v:E\triangleright \kappa: \Gamma, M: ?C$}
\AxiomC{$x:C\triangleright \overline{P}\ |\ \overline{N}: ?\Delta, N_0: ?A, N_1: ?A$}
\BinaryInfC{$v:E\triangleright \kappa: \Gamma, \overline{P}[x:= \mathtt{x}(M)],\hskip1in$} 
\noLine
\UnaryInfC{$\bigl[\mathtt{store}(\langle \overline{N}, [N_0, N_1]\rangle, 
\langle\overline{\mathtt{y}}, \mathtt{y}\rangle, \mathtt{x},M)\ |\ 
\overline{\mathtt{y}}(\mathtt{x}(M)):\ ?\Delta, \mathtt{y}(\mathtt{x}(M)):\, ?A = \bigr.$}
\noLine
\UnaryInfC{$\bigl. = \mathtt{store}(\langle \overline{N}, N_0, N_1\rangle, 
\langle\overline{\mathtt{y}}, \mathtt{y}_0, \mathtt{y}_1\rangle, \mathtt{x},M)\ |\ 
\overline{\mathtt{y}}(\mathtt{x}(M)):\ ?\Delta,
[\mathtt{y}_0(\mathtt{x}(M)), \mathtt{y}_1(\mathtt{x}(M))] :\, ?A \bigr]$}
\DisplayProof\\
\\
{\bf Monoid 1}\\
\AxiomC{$v:E\triangleright \kappa: \Gamma, M:\, ?C$}
\AxiomC{$R \in \kappa \cup M$}
\BinaryInfC{$v:E\triangleright \kappa: \Gamma, \bigl[\,[M, \mathtt{connect\ to}(R)]:\, ?C = M:\, ?C\, \bigr]$}
\DisplayProof\\
\\
{\bf Monoid 2}\\
\AxiomC{$v:E\triangleright \kappa: \Gamma, M:\, ?C$}
\AxiomC{$R \in \kappa \cup M$}
\BinaryInfC{$v:E\triangleright \kappa: \Gamma, \bigl[\,[\mathtt{connect\ to}(R), M]:\, ?C = M:\, ?C\,\bigr]$}
\DisplayProof\\
\\
{\bf Monoid 3}\\
\AxiomC{$v:E\triangleright \kappa: \Gamma, M_0:\, ?C, M_1:\, ?C$}
\UnaryInfC{$v:E\triangleright \kappa: \Gamma, \bigl[\,[M_0, M_1]:\, ?C = [M_1, M_0]:\, ?C\,\bigr]$}
\DisplayProof\\
\\
{\bf Monoid 4}\\
\AxiomC{$v:E\triangleright \kappa: \Gamma, M_0:\, ?C, M_1:\, ?C, M_2:\, ?C$}
\UnaryInfC{$v:E\triangleright \kappa: \Gamma, \bigl[\,[[M_0, M_1], M_2]:\, ?C = [M_0, [M_1, M_2]]:\, ?C\,\bigr]$}
\DisplayProof\\
\\
\hline
  \end{tabular} 
  \end{small}
  \caption{Categorical Equations in Context}
  \label{categorical-equincontext}
 \end{center} 
\end{table}

\medskip

\noindent
The key decision, discussed at length in G.~M.~Bierman's thesis \cite{Bierman94} pp.~127-131, 
arises in the analysis of the {\em Dereliction-Storage} reduction given by the equation in context in Table
\ref{storage-reductions}. By repeating for the rules of {\em dereliction} and {\em storage} the kind of analysis done for {\em par}, {\em subtraction} and {\em unit},  we see that in order to model the 
{\em storage} rule we need a natural transformation 
$\Phi_{E, \Gamma}: \mathbb{C}(E, \Gamma\bullet ?A) \times \mathbb{C}(A, ?\Delta) \rightarrow 
\mathbb{C}(E, \Gamma\bullet ?\Delta)$. By naturality considerations this is given by its action 
$\Phi_{\Gamma}(id_{\Gamma\bullet ?A}, d) =_{df} d^*$ on morphisms $d: A\rightarrow ?\Delta$. 
Similarly, for the {\em dereliction} rule we need a natural transformation 
$\Psi:  \mathbb{C}(\_, A) \rightarrow \mathbb{C}(\_, ?A)$ and by applying Yoneda's Lemma we see
that its action is given by a morphism $\eta_A: A \rightarrow ?A$.

\medskip

\noindent
We can certainly define a functor $\mathbf{?}: \mathbb{C}(A, \Gamma) \rightarrow \mathbb{C}(?A, ?\Gamma)$ by $f \mapsto (f; \eta_{\Gamma})^*$. Now by the equation in context for {\em dereliction}-{\em storage} we have that following the diagram commutes: 
\begin{center}
\begin{tabular}{c}  
$\xymatrix@R=.4in{
??A & ??A \ar[l]_{(\eta_A)^*} \\
& ?A \ar[ul]^{?\eta_A} \ar[u]_{\eta_{?A}}\\
}$
\end{tabular}
\end{center}
Assuming the above decomposition to be unique, we have $(\eta_A)^* = id_{??A}$ and thus the derivations 
$$\eta_{?A}:\  x:?A \triangleright [x]: ??A\quad \hbox{and}\quad ?\eta_A:\  
z:?A\triangleright \mathtt{store}([\,[x]\,], \mathtt{y}, \mathtt{x}, z)\ |\ \mathtt{y}(\mathtt{x}(z)):??A$$
must be identified. Now it can be shown that identifying $\eta_{?A}$ and $?\eta_A$ forces the functor {\bf ?} to be {\em idempotent}: $\mathbf{??} f = \mathbf{?} f$. In order to avoid such collapse, the functor {\bf ?} is only assumed to be a {\bf K} modality, and the properties of {\bf S4} are given by the natural transformations $\eta: A\rightarrow ?A$ and $\mu: ??A\rightarrow ?A$ of the {\em monad} 
$(\mathbf{?}, \eta, \mu)$. Here $\mu_A$ is given by the proof 
\[z: ??A\triangleright \mathtt{store}(x, \mathtt{y}, \mathtt{x}, z)\ |\ \mathtt{y}(\mathtt{x}(z)) : ?A
\]
and the commutative diagram required by the definition of a monad  
\begin{center}
\begin{tabular}{c}  
$\xymatrix@R=.4in{
?A &\\
??A \ar[u]^{\mu_A}&?A \ar[l]_{?\eta_A} \ar[ul]_{id_{?A}}\\
}$
\end{tabular}
\end{center}
identifies $id_{?A}: x: ?A \triangleright x:?A$ with the following derivation $?\eta_A ; \mu_A$: 
\begin{small}
\begin{center}
\begin{tabular}{c}
\AxiomC{$?\eta_A:$}
\noLine
\UnaryInfC{$\quad x:A\triangleright x: A$}
\UnaryInfC{$\qquad x:A\triangleright [x]: ?A$}
\UnaryInfC{$z: ?A \triangleright z: ?A\qquad x:A\triangleright [\,[x]\,]: ??A$}
\UnaryInfC{$z:?A\triangleright \mathtt{store}([\,[x]\,], \mathtt{y}, \mathtt{x}, z)\ |\ t: ??A$}
\AxiomC{$\mu_A:$}
\noLine
\UnaryInfC{$z':??A \triangleright z':??A\quad x':?A \triangleright x':?A$}
\UnaryInfC{$z': ??A\triangleright \mathtt{store}(x', \mathtt{y}', \mathtt{x}', z')\ |\ 
\mathtt{y}'(\mathtt{x}'(t)): ?A$}
\BinaryInfC{$z:?A\triangleright \mathtt{store}([\,[x]\,], \mathtt{y}, \mathtt{x}, z), 
\mathtt{store}(x', \mathtt{y}', \mathtt{x'}, t)\ |\ \mathtt{y}'(\mathtt{x}'(t)):\, ?A$}
\noLine
\UnaryInfC{where $t = \mathtt{y}(\mathtt{x}(z)):??A$}
\DisplayProof\\
\end{tabular}
\end{center}
\end{small}
The normal form of the derivation $?\eta_A ; \mu_A$ is the following one:
\[
z:?A\triangleright \mathtt{store}([\,[x]\,], \mathtt{y}, \mathtt{x}, z)\ |\ \mathtt{y}'(\mathtt{x}'(z)):\, ?A
\]
as in the Categorical Equation in Context for Monad of Table \ref{categorical-equincontext}. 
Further details are left to the reader.
\section{Conclusion.}
In order to provide a categorical semantics for {\em co-intuitionistic logic} - given that as remarked by Tristan Crolard \cite{Crol01} co-exponents in the category {\bf Set} are trivial - we have given a categorical semantics for {\em intuitionistic multiplicative and exponential} co-intuitionistic linear logic, from which our desired results follows by dualizing J-Y.~Girard's embedding of intuitionistic logic into intuitionistic linear logic. 

\medskip

\noindent
In this task we started from a term assignment to multiplicative co-intuitionistic logic, which has been proposed as an abstract {\em distributed calculus} dualizing the linear $\lambda$ calculus 
\cite{BellinLAM,Bellin05,BellinMenti}: in our view such dualization underlies the translation of the linear 
$\lambda$-calculus into the $\pi$-calculus (see \cite{BellinScott}). Our {\em dual distributed calculus} is itself a restriction to a co-intuitionistic consequence relation of Crolard's term assignment to 
{\em subtraction} in the framework of the $\lambda\mu$-calculus: to {\em subtraction introduction} 
and {\em elimination} rules and to their $\beta$ reduction {\em global operations of binding} and {\em global substitution} are assigned; these operations may appear as notationally awkward at first sight 
but are forced on us by the removal of the $\mu$-rule and of the $\mu$-variable abstraction used in Crolard's approach. A computational application of our notation is suggested in the proof of the 
Decomposition Property (Proposition \ref{Decomposition}). Since the dependencies of a variable $y : C$ 
from $n$ binders $\mathtt{make}-\mathtt{coroutine}$ and $\mathtt{postpone}$ are represented in a 
term $\mathtt{y}(t_1(\ldots t_n(M)\ldots)) : C$ by the terms $t_i : D_i$, this representation can be used to  compute the assignment of a probabilistic event $\mathbf{C}$ to such a term according to the assignments $\mathbf{D}_i$ to the terms  $t_i : D_i$ and to prove that \emph{probabilities are
preserved} from the premise to the disjunction of the conclusions in a multiplicative co-intuitionistic 
derivation.  

\medskip

\noindent
Our work required a lengthy exercise on well-known results by Benton, Bierman, Hyland and 
de Paiva\cite{BBHdP92,Bierman94}, with the considerable help given by Blute, Cockett, Seely and Trimble's work \cite{BCS,BCST}. To assess the merits and advantages of our work we need to evaluate 
the syntax for the exponential rules: here again the {\em storage} rule may appear notationally quite heavy, but it is a straightforward  implementation of the act of storing. 
On the other hand the advantages of working in the dual system are completely evident in the treatment of {\em dereliction} and {\em contraction}, where the awkward {\tt let} operations and related naturality conditions are replaced by simple operations on lists. Finally, the treatment of {\em weakening} is also completely standard, thanks also to Blute, Cockett,  Seely and Trimble's work \cite{BCS,BCST} on the notion of {\em rewiring}.
\bibliographystyle{alpha}

\appendix
\section{Example} \label{example}

Consider  the following computation in the simply typed $\lambda$-calculus. We write $\mathbf{N}$
for $(A \supset A) \supset (A \supset A)$ and  $\mathbf{N}^{\bot}$ for 
$(C \smallsetminus C) \smallsetminus (C \smallsetminus C)$.
\begin{center}
{\small
 \begin{tabular}{rcc}
$\lambda h^{A\supset A}.(\lambda f^{A\supset A}.\lambda x^A.ffx)
(\lambda g^{A\supset A}\lambda y^A.ggy) h : \mathbf{N}$&
$\rightsquigarrow_{\beta}$&\qquad (i)\\
$\lambda h^{A\supset A}.(\lambda f^{A\supset A}.\lambda x^A.ffx)(\lambda y^A.hhy) : \mathbf{N}$&  
$\rightsquigarrow_{\beta}$&\qquad (ii)\\
$\lambda h^{A\supset A}.\lambda x^A.(\lambda y^A.hhy)(\lambda y^A.hhy)x : \mathbf{N}$&  
$\rightsquigarrow_{\beta}$&\qquad (iii)\\
$\lambda h^{A\supset A}.\lambda x^A.((\lambda y^A.hhy)hhx: \mathbf{N}$&
$\rightsquigarrow_{\beta}$&\qquad (iv)\\
$\lambda h^{A\supset A}.\lambda x^A.hhhhx : \mathbf{N}$& &\qquad  $(v)$\\
\end{tabular} }
\end{center}
 
\begin{table}[ht]
 \begin{center}
  \begin{small}
 \begin{displaymath}
 \begin{array}{l}
 \xymatrix@R=.2in{\ar@{=>}[ddddddddddd]& & & & & & & & & & \\
& & (2)\ar@{-}[dr]|{f : A \rightarrow A} & &(1)\ar@{-}[dl]|{x : A} & \\
& (2)\ar@{-}[dr]|{f : A\rightarrow A}& &\ar@{-}[dl]|{fx : A} & \\
& & \ar@{-}[d]|{ffx : A} & & & &  \\
& & {(1)}\ar@{-}[d]|{\lambda x.ffx:A\rightarrow A}&&(4)\ar@{-}[dr]|{h:A\rightarrow A}&&(3)\ar@{-}[dl]|{y: A}\\    
& &(2)\ar@{-}[dddr]|{\underbrace{\lambda f \lambda x.ffx}_{\mathbf{t}} : \mathbf{N}}&(4)\ar@{-}[dr]|{h : A\rightarrow A} & &\ar@{-}[dl]|{hy: A} & &\\
& & & &\ar@{-}[d]|{hhy:A} & & & \\
& & & & {(3)}\ar@{-}[dl]^{\underbrace{\lambda y.hhy}_{\mathbf{u}} : A\rightarrow A } & & \\
& & & \ar@{-}[d]|{\mathbf{tu} : A \rightarrow A} & & \\
& & & {(4)}\ar@{-}[dd]|{\underbrace{\lambda h.(\lambda f \lambda x.ffx)(\lambda y.hhy)}_{\lambda h.\mathbf{tu}}: \mathbf{N}}  \\
& & & & &  & &\\
& & & & &  & &}
\end{array}
\end{displaymath}
  \end{small}
  \caption{Natural Deduction tree for  
  (ii)\ $\vdash \lambda h.(\lambda f.\lambda x.ffx)(\lambda y.hhy) : \mathbf{N}$}
  \label{two-times-two}
 \end{center} 
\end{table}

\begin{table}[ht]
 \begin{center}
  \begin{small}
 \begin{displaymath}
 \begin{array}{l}
 (2): P_2 = \underbrace{\mathtt{postp}(g\mapsto [m_2,m_1], m_3)}_{\mathbf{Redex} }\quad 
 (4): P_4 = \mathtt{postp}(e\mapsto [m_5,m_4], n)\\
 (1): P_1 = \mathtt{postp}(b\mapsto [x], g)\qquad\qquad (3): P_3 =  \mathtt{postp}(d\mapsto [y], j)\\
 \xymatrix@R=.2in{& & & & & & & & & & \\
& & (2)\ar@{-}[dr]|{m_1= \mathtt{mkc}(a,\mathtt{x}): C\smallsetminus C} & &(1)\ar@{-}[dl]^{x = \mathbf{x}(a): C} & \\
& (2)\ar@{-}[dr]|{m_2= \mathtt{mkc}(b, \mathtt{a}): C\smallsetminus C}& &\ar@{-}[dl]^{a = \mathbf{a}(b): C} & \\
& & \ar@{-}[d]|{b = \mathtt{b}(g): C} & & & & & & & &  \\
& & {(1)}\ar@{-}[d]|{g = \mathtt{g}(m_3):C\smallsetminus C}&&(4)\ar@{-}[dr]|{m_4 = \mathtt{mkc}(c,\mathtt{y}):C\smallsetminus C}&&(3)\ar@{-}[dl]^{y = \mathtt{y}(c): C}\\    
& &(2)\ar@{-}[dddr]|*+{\underline{\mathbf{Red}}\ m_3 = \mathtt{mkc}(e, \mathtt{j}): \mathbf{N}^{\bot}}&(4)\ar@{-}[dr]|{m_5 = \mathtt{mkc}(d,\mathtt{c}) : C\smallsetminus C} & &\ar@{-}[dl]^{c = \mathtt{c}(d) : C} & & & & \\
& & & &\ar@{-}[d]^{d = \mathtt{d}(j):C} & & & & &\\
& & & & {(3)}\ar@{-}[dl]|{j = \mathtt{j}(e) : C\smallsetminus C } & & & & & &   \\
& & & \ar@{-}[d]|{e = \mathtt{e}(n):C\smallsetminus C} & & & & & &  \\
& & & (4)\ar@{-}[d]|{n: \mathbf{N}^{\bot}} & & & &   \\
\ar@{=>}[uuuuuuuuu]& & & &   }
\end{array}
\end{displaymath}
  \end{small}
  \caption{{\bf co-IL} tree for the dual of (ii)\ $\vdash \lambda h.(\lambda f.\lambda x.ffx)(\lambda y.hhy) : \mathbf{N}$}
  \label{co-two-times-two}
 \end{center} 
\end{table}

\begin{table}[ht]
 \begin{center}
  \begin{small}
 \begin{displaymath}
 \begin{array}{l}
 (1): P'_1 = \mathtt{postp}(b\mapsto [x], e)\qquad
 \qquad(4): P'_4 = \mathtt{postp}(e\mapsto[m'_5,m''_5,m'_4,m''_4], n)\\ 
 (3'): P'_3 = \underbrace{\mathtt{postp}(d'\mapsto [y'], m_2)}_{\mathbf{Redex}}\qquad
 (3''): P''_3 = \underbrace{\mathtt{postp}(d''\mapsto [y''], m_1)}_{\mathbf{Redex}}\\
 \xymatrix@R=.2in{& & & & & & & & & & \\
& & & & &  (4)\ar@{-}[dr]|{m''_4 = \mathtt{mkc}(c'',\mathtt{y}):C\smallsetminus C} %
  & &(3'')\ar@{-}[dl]^{y'' = \mathtt{y}(c''): C}&\\
& & & & (4)\ar@{-}[dr]|{m_5'' =\mathtt{mkc}(d'',\mathtt{c}) : C\smallsetminus C}&&%
\ar@{-}[dl]^{c'' = \mathtt{c}(d''):C} & \\
& & & & &    \ar@{-}[dd]|{d'' = \mathtt{d}(m_1):C} & \\
& & (4)\ar@{-}[dr]|{m'_4 = \mathtt{mkc}(c',\mathtt{y}):C\smallsetminus C} %
  & &(3')\ar@{-}[dl]^{y' = \mathtt{y}(c'): C}& \\
&(4)\ar@{-}[dr]|{m'_5 =\mathtt{mkc}(d',\mathtt{c}) : C\smallsetminus C}&&%
\ar@{-}[dl]^{c' = \mathtt{c}(d'):C} & & %
(3'') \ar@{-}[dr]|*+{\underline{\mathbf{Red}''}m_1= \mathtt{mkc}(a,\mathtt{x}): C\smallsetminus C}%
&&&  (1)\ar@{-}[dll]^{x = \mathbf{x}(a): C} & \\
& &\ar@{-}[d]|{d' = \mathtt{d}(m_2):C}  & & & & \ar@{-}[ddlll]^{a = \mathbf{a}(b): C} & \\
& &(3')\ar@{-}[dr]|*+{\underline{\mathbf{Red}'}\ m_2= \mathtt{mkc}(b, \mathtt{a}): C\smallsetminus C}\\ 
& & & \ar@{-}[d]|{b = \mathtt{b}(e): C} & & & \\
& & &(1) \ar@{-}[d]|{g = e = \mathtt{e}(n):C\smallsetminus C} & & &  \\
& & & (4)\ar@{-}[d]|{n: \mathbf{N}^{\bot}} & & &   \\
\ar@{=>}[uuuuuuuuuu]& & &  \\ }
\end{array}
\end{displaymath}
  \end{small}
  \caption{{\bf co-IL} tree for the dual of 
  (iii)\ $\vdash \lambda h.\lambda x.(\lambda y.hhy)(\lambda y.hhy)x : \mathbf{N}$}
  \label{co-two-times-two-red}
 \end{center} 
\end{table}

\begin{table}[ht]
 \begin{center}
  \begin{small}
 \begin{displaymath}
 \begin{array}{l}
(2): P_2 = \underbrace{\mathtt{postp}(g\mapsto [[m_1][m_2]], m_3)}_{\mathbf{Redex} }\\
(1): P_1 = \mathtt{postp}(b\mapsto x, g)\qquad\qquad\qquad %
(4'): P_4 = \mathtt{postp}(e\mapsto [y_1], n)\\
\xymatrix@R=.2in{& & & & & & & & & & \\
&(2)\ar@{-}[dd]|{\underbrace{[[m_1][m_2]]: ?(C\smallsetminus C)}_{M_0}} & &\\
&     &     & \\
& \ar@{-}[d]|{[m_2]: ?(C\smallsetminus C)} \ar@{-}[r]^{[m_2]: ?(C\smallsetminus C)} %
 &\ar@{-}[dr]|{m_1 = \mathtt{mkc}(a,\mathtt{x})}& &(1)\ar@{-}[dl]|{x = \mathtt{x}(a)} & %
 (4') \ar@{-}[d]|{[y_0]: ?(C\smallsetminus C)}& \\
& \ar@{-}[dr]|{m_2 = \mathtt{mkc}(b,\mathtt{a})}& & \ar@{-}[dl]|{a = \mathtt{a}(b)} & & & & \\
& &\ar@{-}[dd]|{b = \mathtt{b}(g): C} %
  & & \ar@{--}[rrrr] &\ar@{-}[u]|*+{y_0 = \mathtt{y}_0(\mathtt{k}(j))} & & & & & \\
& & & & &(4)\ar@{-}[u]|{M = [[m_5][m_4]]}\ar@{-}[d]|{M : ?(C\smallsetminus C)} &\mathbf{B}_{k,M}& & &\\
& &(1)\ar@{-}[ddd]|{g = \mathtt{g}(m_3): C\smallsetminus C} %
 & & & \ar@{-}[d]|{[m_5] : ?(C\smallsetminus C)} &\ar@{-}[l]_{[m_4] : ?(C\smallsetminus C)} %
                 \ar@{-}[dr]|{m_4= \mathtt{mkc}(c, \mathtt{y}): C\smallsetminus C} & &(3) \qquad& & \\
& & & & &\ar@{-}[dr]|{m_5= \mathtt{mkc}(d, \mathtt{c}): C\smallsetminus C}& & %
   \ar@{-}[dl]^{c= \mathtt{c}(d) :C} \ar@{-}[ur]_{y = \mathtt{y}(c) :C}& & &\\  
& & & & & & \ar@{-}[d]|{d = \mathtt{d}(k):C} & & & & & \\
& &(2)\ar@{-}[dddr]|*+{\underline{\mathtt{Red}}\ m_3 = \mathtt{mkc}(e, \mathtt{j}): \mathtt{N}^{\bot}} %
 & & & & (3)\ar@{-}[d]|{k: C\smallsetminus C} & & & & &\\
& & & & & & %
\ar@{-}[d]_*+{\mathtt{store}(P_3, M, \mathtt{y}_0, \mathtt k, j)}^*+{(3): P_3=\mathtt{postp}(d\mapsto y, j)} %
& &  & \\ 
& & & & \ar@{--}[uuuuuuu]\ar@{--}[rrrr]|*+{k = \mathtt{k}(j)}& & 
   \ar@{-}[dlll]|{j = \mathtt{j}(e) : ?(C\smallsetminus C)} & & \ar@{--}[uuuuuuu] &   \\
& & & \ar@{-}[d]|{e = \mathtt{e}(n): C\smallsetminus C} & & & &  \\
& & & (4')\ar@{-}[d]|{n: \mathbf{N}^{\bot}} & & & &   \\
\ar@{=>}[uuuuuuuuuuuu]& & & &   }
\end{array}
\end{displaymath}
  \end{small}
  \caption{{\bf linear co-IL} analysis of the proof in Table \ref{co-two-times-two}}
  \label{lin-co-two-times-two}
 \end{center} 
\end{table}

\begin{table}[ht]
 \begin{center}
  \begin{small}
 \begin{displaymath}
 \begin{array}{l}
 (1): P_1 = \mathtt{postp}(b\mapsto x, e)\qquad\qquad\qquad %
(4'): P_4 = \mathtt{postp}(e\mapsto [[y_2][y_1]], n)\\
\xymatrix@R=.2in{& & & & & & & & & & \\
& & &(4')\ar@{-}[d]|*+{[[y_2][y_1]]} &\\
& & &\ar@{-}[dll]|*+{[y_2]}\ar@{-}[dr]|*+{[y_1]} &\\
&\ar@{-}[d]|*+{y_2 =\mathtt{y}_0(k_2)}& & & \ar@{-}[d]|*+{y_2 =\mathtt{y}_0(k_2)}&\\          
\ar@{--}[rr]&                   & &\ar@{--}[rr] &                      & \\ 
& \mathbf{B}_{k_2,M_2}& & &\mathbf{B}_{k_1,M_1} & \\
\ar@{--}[uu]\ar@{--}[rr]& &\ar@{--}[uu] &\ar@{--}[uu]\ar@{--}[rr]&  &\ar@{--}[uu] \\ 
&\ar@{-}[d]|{[m_2] : ?(C\smallsetminus C)}\ar@{-}[uu]|*+{k_2=\mathtt{k}([m_2])}& %
       & & \ar@{-}[d]|{[m_1] : ?(C\smallsetminus C)} \ar@{-}[uu]|*+{k_1=\mathtt{k}([m_1])}&\\ 
&\ar@{-}[ddrr]|{m_2 = \mathtt{mkc}(b, \mathtt{a}}& & &\ar@{-}[dr]|{m_1 = \mathtt{mkc}(a,\mathtt{x})} 
           & &(1)\ar@{-}[dl]|{x = \mathtt{x}(a)} &\\
& & &                          &              &\ar@{-}[dll]|{a = \mathtt{a}(b):C} & \\ 
& & &\ar@{-}[d]|{b = \mathtt{b}(g): C} & & & \\
& & &(1) \ar@{-}[d]|{g = e = \mathtt{e}(n): C\smallsetminus C} & & & &  \\
& & &(4') \ar@{-}[d]|{n: \mathbf{N}^{\bot}} & & & &   \\
\ar@{=>}[uuuuuu]& & & &   }
\end{array}
\end{displaymath}
  \end{small}
  \caption{{\bf linear co-IL} analysis of the proof in Table \ref{co-two-times-two-red}}
  \label{lin-co-two-times-two-red}
 \end{center} 
\end{table}

\noindent
In Table \ref{two-times-two} we give a Natural Deduction derivation in ``tree form'' with the assignment 
of the term (ii)\ $\vdash \lambda h.(\lambda f.\lambda x.ffx)(\lambda y.hhy) : \mathbf{N}$. In Table 
\ref{co-two-times-two} we consider a Natural Deduction derivation 
of $n : \mathbf{N}^{\bot}\vdash P_1, P_2, P_3, P_4\; |$ in the \emph{subtraction only fragment of 
co-intuitionistic logic}; such a derivation is exactly dual of that in Table \ref{two-times-two}: it is also 
in ``tree form'', in fact it yields the same tree as in Table \ref{two-times-two} read from bottom up. 
Its term assignment, with the p-terms $P_1, P_2, P_3, P_4$ in the conclusion, belong to a
\emph{dual calculus} defined in \cite{Bellin12} and briefly described here, with the property that to each 
$\beta$ reduction in the simply typed $\lambda$ calculus there corresponds a set of rewritings of the 
computational context in the dual calculus and a reduction sequence $t_1, t_2, \ldots$ of simply typed 
$\lambda$ terms terminates if and only if the dual sequence $t^{\bot}_1, t^{\bot}_2, \ldots$ terminates
(see \cite{Bellin12}, section 6.1).
The derivation in Table \ref{co-two-times-two-red} results from that in Table \ref{co-two-times-two} 
by one step of normalization. 

\vspace{1ex}
\goodbreak

\noindent
The grammar of the \emph{dual calculus} for the subtraction-only fragment of co-intuitionism is as follows (see \cite{Bellin12}, section 6, definition 10): 
\begin{center}
\begin{tabular}{rl}
$M\ :=$& $ x\ |\ \mathtt{x}(M) \ |\ \hbox{\tt mkc}(M, \mathtt{x})$\\
$\ell\ :=$& $\ []\  |\ [M_1, \ldots, M_n]$ for some $n$.\\
$P\ :=$& $\mathtt{postp}(\mathtt{y}\mapsto \ell[y:= \mathtt{y}(M)], M) $.\\
\end{tabular}
\end{center}
Here non-empty lists $\ell$ are \emph{flattened} and occur only within p-terms. A notion of \emph{term expansion} allows to define the substitution of a \emph{flat list} for a free variable in (a \emph{flat list} of) terms, yielding a \emph{flat list}. In this way we take care of \emph{contraction of discharged conclusions} resulting from a \emph{subtraction elimination}. A conclusion introduced by 
\emph{weakening} is assigned an empty list. To decorate natural deduction trees we use a notation 
which ignores the distinction between local and remote binding, as discussed in Note 
\ref{local-remote-binding} (iii). Notice that the grammar of the \emph{dual calculus} used in this 
section is actually a fragment of the grammar of the \emph{linear dual calculus} presented in this paper
\footnote{We have not explored the possibility of assigning the \emph{empty list} to the formulas $? C$ introduced by \emph{weakening} also in the linear calculus; this would distinguish the case of \emph{weakening} from that of the $\bot$-\emph{introduction} rule, where terms of the form 
$\mathtt{connect\ to}\;(R)$ would still be used.}.

\vspace{3ex}

\noindent
Next we translate the \emph{co-intuitionistic} natural deduction derivations of Tables  
\ref{co-two-times-two} and \ref{co-two-times-two-red} into \emph{co-intuitionistic linear} logic. 
In Tables \ref{lin-co-two-times-two} and \ref{lin-co-two-times-two-red} we adapt the graphical notation 
of Tables \ref{co-two-times-two} and \ref{co-two-times-two-red} to our linear calculus and notice
that that the derivation in Table \ref{lin-co-two-times-two-red} results from that in 
Table \ref{lin-co-two-times-two} by applying \emph{two normalization steps}, a \emph{subtraction 
reduction} followed by a \emph{storage - contraction reduction}. The graphical notation should help 
to catch a glimpse of the reduction process more vividly. Here we present the Natural Deduction 
derivation of Table \ref{lin-co-two-times-two} in the \emph{sequent-style typing judgements} of our official 
calculus. 

\vspace{3ex}

\noindent
{\bf Sequent-style Natural Deduction.} (i)\ The derivation $\mathcal{D}$ corresponding to the graph inside \emph{box} $\mathbf{B}_{k,M}$ of Table \ref{lin-co-two-times-two} is as follows.
\begin{small}
\begin{center}
\AxiomC{$k:C\smallsetminus C\triangleright\ |\ k:C\smallsetminus C$}
\AxiomC{$d:C\triangleright\ |\ d:C$}
\AxiomC{$c:C\triangleright\ |\ c:C$}
\LeftLabel{$_{\smallsetminus I}$}
\BinaryInfC{$d:C\triangleright\ |\ \underbrace{\mathtt{mkc}(d,\mathtt{c})}_{m_5}: C\smallsetminus C, %
\underbrace{\mathtt{c}(d)}_c:C $}
\AxiomC{$y:C\triangleright\ |\ y:C$}
\LeftLabel{$_{\smallsetminus I}$}
\BinaryInfC{$d:C\triangleright\ |\ m_5: C\smallsetminus C, 
\underbrace{\mathtt{mkc}(c,\mathtt{y})}_{m_4}: C\smallsetminus C, \underbrace{\mathtt{y}(c)}_y :C $}
\LeftLabel{$_{der}$}
\UnaryInfC{$d:C\triangleright\ |\ [m_5] : ?(C\smallsetminus C), [m_4]: ?(C\smallsetminus C), y:C$}
\LeftLabel{$_{contr}$}
\UnaryInfC{$d:C\smallsetminus C\triangleright\ |\ [[m_5][m_4]]: ?(C\smallsetminus C), y:C$}
\LeftLabel{$_{\smallsetminus E}$}
\BinaryInfC{$k:C\smallsetminus C\triangleright \underbrace{\mathtt{postp}(d\mapsto y, k)}_{P_3}\ |\ %
\underbrace{[[m_5][m_4]] [d := \mathtt{d}(k)]}_M : ?(C\smallsetminus C)$}
\DisplayProof
\end{center}
\end{small}
Applying the $?$-E rule with major premise 
$j: ?(C\smallsetminus C) \triangleright j: ?(C \smallsetminus C)$ we obtain a derivation 
of the following sequent:  
\[
j: ?(C\smallsetminus C)\ \triangleright\ \mathtt{store}(P_3, M, \mathtt{y}_0, \mathtt{k}, j)\ |\ 
\mathtt{y}_0(\mathtt{k}(j)): ?(C\smallsetminus C).
\]
Finally, by applying \emph{subtraction introduction} to it with the axiom \newline
$e: C\smallsetminus C \triangleright \ |\ e: C\smallsetminus C$ we obtain a derivation 
$\mathcal{D}_1$ of the following sequent:
\[
e: C\smallsetminus C\ \triangleright\ 
\underbrace{\mathtt{store}(P_3, M, \mathtt{y}_0, \mathtt{k}, \mathtt{j}(e))}_{\mathbf{Store}}\ |\ 
\underbrace{\mathtt{y}_0(\mathtt{k}(\mathtt{j}(e)))}_{y_0} : ?(C\smallsetminus C), \mathtt{mkc}(e, \mathtt{j}) : 
\mathbf{N}^{\bot}
\]
where $\mathbf{N}^{\bot} = (C\smallsetminus C) \smallsetminus ?(C\smallsetminus C)$. 

\vspace{1ex}

\noindent
(ii) By applying the same steps as in derivation $\mathcal{D}$, but relabelling of the terms, we obtain 
a derivation $\mathcal{D}_0$ of the following sequent: 
\[
g:C\smallsetminus C\triangleright \underbrace{\mathtt{postp}(b\mapsto x, g)}_{P_1}\ |\ %
\underbrace{[[m_1][m_2]] [b := \mathtt{b}(g)]}_{M_0} : ?(C\smallsetminus C)
\]
where $m_1 = \mathtt{mkc}(b, \mathtt{a})$, $m_2 = \mathtt{mkc}(a, \mathtt{x})$, 
$b = \mathtt{b}(g)$, $a = \mathtt{a}(b)$ and $x = \mathtt{x}(a)$.

\noindent
(iii) Now if we apply $\mathcal{D}_1$ to $\mathcal{D}_0$ with the formula $\mathtt{mkc}(e, \mathtt{j}) : 
\mathbf{N}^{\bot}$ as major premise of \emph{subtraction elimination} then we obtain  
a derivation $\mathcal{D}^+$ ending with following sequent:
\[
e: C\smallsetminus C\ \triangleright\ \mathbf{Store}\ 
\mathtt{postp}(g\mapsto M_0, \mathtt{mkc}(e, \mathtt{j}))\ |\ 
y_0: ?(C\smallsetminus C)
\]   
The pair  \emph{introduction / elimination} inferences determines the only {\bf Redex} in $\mathcal{D}^+$. 
A final \emph{subtraction elimination} with the axiom 
$n: \mathbf{N}^{\bot} \triangleright n: \mathbf{N}^{\bot}$ concludes the derivation in our example.

\end{document}